\documentclass[sigconf,nonacm]{acmart}
\usepackage{makecell}
\usepackage{endnotes}
\usepackage{xcolor}
\usepackage{graphicx}
\usepackage{wrapfig}
\usepackage{multirow}
\usepackage{listings}
\usepackage{url}

\setcopyright{none}
\renewcommand\footnotetextcopyrightpermission[1]{}

\usepackage{balance}
\usepackage{listings}
\usepackage{verbatim}
\usepackage{subfigure}
\usepackage{stfloats}
\usepackage{amsmath}
\usepackage{balance}
\usepackage[detect-all,group-separator={,}]{siunitx}
\usepackage{enumitem}

\def\BibTeX{{\rm B\kern-.05em{\sc i\kern-.025em b}\kern-.08em
    T\kern-.1667em\lower.7ex\hbox{E}\kern-.125emX}}

\definecolor{mGreen}{rgb}{0,0.6,0}
\definecolor{mGray}{rgb}{0.5,0.5,0.5}
\definecolor{mPurple}{rgb}{0.58,0,0.82}
\definecolor{backgroundColour}{rgb}{0.95,0.95,0.92}

\lstdefinestyle{CStyle}{
    backgroundcolor=\color{backgroundColour},   
    commentstyle=\color{mGreen},
    keywordstyle=\color{magenta},
    numberstyle=\tiny\color{mGray},
    stringstyle=\color{mPurple},
    basicstyle=\ttfamily\footnotesize,
    breakatwhitespace=false,         
    breaklines=true,                 
    captionpos=b,                    
    keepspaces=true,                 
    numbers=left,                    
    numbersep=5pt,                  
    showspaces=false,                
    showstringspaces=false,
    showtabs=false,                  
    tabsize=2,
    language=C
}

\usepackage{booktabs}

\begin{document}

\pagestyle{plain}

\title{Understanding the Landscape of Ampere GPU Memory Errors}

\author{Zhu Zhu}
\affiliation{%
  \institution{George Mason University}
 \country{USA}}
\email{zzhu22@gmu.edu}

\author{Yu Sun}
\affiliation{%
  \institution{George Mason University}
 \country{USA}}
\email{ysun23@gmu.edu}

\author{Dhatri Parakal}
\affiliation{%
 \institution{University of Illinois Urbana-Champaign}
 \country{USA}}
\email{dhatrip2@illinois.edu}

\author{Bo Fang}
\affiliation{%
 \institution{Pacific Northwest National Laboratory / University of Texas, Arlington}
 \country{USA}}
\email{bo.fang@pnnl.gov}

\author{Steven Farrell}
\affiliation{%
 \institution{National Energy Research Scientific Computing Center}
 \country{USA}}
\email{sfarrell@lbl.gov}

\author{Gregory H. Bauer}
\affiliation{%
 \institution{National Center for Supercomputing Applications}
 \country{USA}}
\email{gbauer@illinois.edu}

\author{Brett Bode}
\affiliation{%
 \institution{National Center for Supercomputing Applications}
 \country{USA}}
\email{brett@illinois.edu}

\author{Ian T. Foster}
\affiliation{%
  \institution{University of Chicago / Argonne National Laboratory}
 \country{USA}}
\email{foster@cs.uchicago.edu}

\author{Michael E. Papka}
\affiliation{%
  \institution{Argonne National Laboratory / 
 University of Illinois Chicago}
 \country{USA}}
\email{papka@anl.gov}

\author{William Gropp}
\affiliation{%
  \institution{University of Illinois Urbana-Champaign}
 \country{USA}}
\email{wgropp@illinois.edu}

\author{Zhao Zhang}
\affiliation{%
 \institution{Rutgers University}
 \country{USA}}
\email{zhao.zhang@rutgers.edu}

\author{Lishan Yang}
\affiliation{%
 \institution{George Mason University}
 \country{USA}}
\email{lyang28@gmu.edu}

 \pagestyle{plain} 

\begin{abstract}

Graphics Processing Units (GPUs) have become a de facto solution for accelerating high-performance computing (HPC) applications. Understanding their memory error behavior is an essential step toward achieving efficient and reliable HPC systems. In this work, we present a large-scale cross-supercomputer study to characterize GPU memory reliability, covering three supercomputers -- Delta, Polaris, and Perlmutter -- all equipped with NVIDIA A100 GPUs. We examine error logs spanning 67.77 million GPU device-hours across 10,693 GPUs. We compare error rates and mean-time-between-errors (MTBE) and highlight both shared and distinct error characteristics among these three systems. Based on these observations and analyses, we discuss the implications and lessons learned, focusing on the reliable operation of supercomputers, the choice of checkpointing interval, and the comparison of reliability characteristics with those of previous-generation GPUs. Our characterization study provides valuable insights into fault-tolerant HPC system design and operation, enabling more efficient execution of HPC applications.

\end{abstract}

\maketitle

\section{Introduction}
~\label{sec:intro}

Graphics Processing Units (GPUs) are widely deployed in modern supercomputers to accelerate high-performance computing (HPC) applications. These HPC systems operate at massive scales, with tens of thousands of GPUs in a single supercomputer.
A range of workloads is supported, from conventional HPC workloads such as fluid simulation~\cite{chen2021gpu,liu2004real} and molecular dynamics~\cite{glaser2015strong,boku2024improving}, to emerging tasks including transformer-based large language models (LLMs)~\cite{radford2018improving, zhang2022opt} and scientific foundation models~\cite{zvyagin2022genslm}.
However, the high-density transistor arrays in GPUs increase their susceptibility to hardware  faults and memory errors  due to cosmic radiation~\cite{DBLP:conf/dsn/FratinOLSRR18} and shrinking transistors~\cite{killi}, leading to silent data corruptions, exceptions, and crashes~\cite{sullivan2021characterizing,opt175logbook,dixit2021silent,he2023understanding,zhang2019quantifying,li2017understanding}.
Even just one single memory error on one GPU can jeopardize the entire application execution, which involve up to $O(10^5)$ of GPUs 
over weeks to months~\cite{opt175logbook,tr176Blogbook}. 
For example, during Meta's OPT-175B training, hardware errors are reported on 18 out of 55 days~\cite{opt175logbook}.
These errors can delay the training process and, in the worst case, compromise model  convergence~\cite{he2023understanding,zhang2019quantifying}.
Therefore, understanding error behavior in large-scale systems is the key first step towards designing error mitigation techniques and avoiding the aforementioned severe consequences of GPU errors.

\vspace{1mm}
\noindent{\bf Limitation of state-of-art approaches.}
Existing supercomputer reliability analyses  do not meet the practical needs for recent large foundation model (LFM) training and scientific applications. 
Many studies focus on CPUs~\cite{levy2018lessons,bautista2016unprotected,ferreira2021understanding,sridharan2013feng} and DRAMs (Dynamic Random Access Memories)~\cite{beigi2023systematic}.
GPU error studies primarily reveal the error characteristics of earlier GPU generations including NVIDIA K20X~\cite{di2014lessons,debardeleben2014gpu,tiwari2015understanding,nie2016large,nie2017characterizing,ostrouchov2020gpu} and V100~\cite{oles2024understanding}.
Moreover, past GPU error characterization studies typically focus on a single supercomputer~\cite{oles2024understanding,nie2017characterizing,tiwari2015understanding,cui2025characterizing}. 
This lack of a cross-supercomputer perspective may result in observations biased by the studied system and thus not generalizable to other HPC systems.

\noindent{\bf Motivation and Challenges.}
To address the growing computing demands for the new LFM paradigm, academia and industrial stakeholders have built new supercomputers with massive numbers of GPUs.
This widespread deployment and growing popularity highlights the necessity of revisiting GPU error characteristics in production supercomputers, with the goal of  deriving effective operation suggestions for HPC users and system administrators to preserve high execution efficiency and machine utilization.

As we demonstrate in the paper, a {\it cross-supercomputer} perspective is essential in such error behavior study. 
While clusters with the same GPU generation have common reliability traits, 
their unique hardware settings and workloads may introduce unknown biases.
Performing a cross-supercomputer study helps remove these biases to uncover the natural behavior of GPU errors.
However, such study is challenging:
1) the differences in supercomputer configurations complicate the analysis;
2) the heterogeneity of error data across systems renders data discovery and fair comparisons difficult.

\vspace{1mm}
\noindent{\bf Key insights and contributions.}
We perform a cross-institute collaboration on a comprehensive error characterization study across \num{10693} NVIDIA Ampere GPUs in three heavily utilized, in-production supercomputers: Delta~\cite{delta}, Polaris~\cite{polaris}, and Perlmutter~\cite{perlmutter}.
We monitor and record ECC-reported errors of GPUs to form the error log, i.e., a GPU memory error dataset, resulting in a total of 67.77 million GPU device-hours.
Our key observations are summarized as follows: 
\begin{itemize}

\item The observed error rates on different clusters can vary by over three orders of magnitude. Trusting the results from one single cluster may introduce significant biases in error behavior characterization.
\item Errors are not uniformly distributed over time.
Burstiness is generally observed in all three clusters, while their severeness varies.
\item Both bursty error patterns and supercomputer scale can affect the characteristics of error interarrival times. Using a single number, the mean-time-between-error (MTBE), may introduce bias.  Instead, the distribution of interarrival times should be considered.
\item There is no observable periodicity of errors.
\item For environmental factors such as temperature, power, and GPU utilization, we do not observe any strong correlation with errors.
\item Erroneous nodes vary over time. 

\end{itemize}

Based on these observations and analyses, we further discuss the implications and lessons learned from this study, focusing on the reliable operation of supercomputers, the choice of checkpointing interval, and the comparison of reliability characteristics with those of previous-generation GPUs.
Our key take-away messages are:
\begin{itemize}
  \item  NVIDIA V100 and A100 GPUs both use
HBM2 and have similar memory error characteristics.
    
\item  Coarser-level error monitoring
does not necessarily lead to much information loss, yet monitoring
errors at a finer level enables faster responses to errors.
\item  Dynamic node monitoring and
erroneous GPU prediction strategies are needed for the
efficient and reliable operation of supercomputers.
\item Dynamic checkpointing is suggested to accommodate the bursty error patterns observed in all three supercomputers. GPU error monitoring could be used to guide the frequency of dynamic checkpointing.

\end{itemize}

\section{Background and Data Collection Method}
~\label{sec:background}

We first describe the architecture and organization of the studied supercomputers and their constituent GPUs. Then, we discuss our data collection method and the basic information contained in our datasets. Last but not least, we discuss the scope and limitations of this work.

\subsection{Supercomputer Organizations}

We focus on  three highly utilized, in-production supercomputers in this error characterization  study.
Below we briefly describe the architecture of each supercomputer.

\vspace{1mm}
\noindent {\bf Delta~\cite{delta}} is a supercomputer
designed by Hewlett Packard Enterprise (HPE) and National Center for Supercomputing Applications (NCSA).
There are three types of NVIDIA GPU nodes, with 849 GPUs in total: Delta has 100 four-way NVIDIA A40 GPU compute nodes, each equipped with a single 2.45 GHz AMD Milan 64-core CPU.
The four A40 GPUs are connected via PCIe.
Another 100 four-way GPU compute nodes have NVIDIA A100s connected via NVLink, with all other configurations identical to the A40 nodes. 
There also exist six eight-way NVIDIA A100 GPU compute nodes, each with two 2.45 GHz AMD Milan 64-core CPU sockets.
The eight A100 GPUs are connected via NVLink.

\vspace{1mm}
\noindent {\bf Polaris~\cite{polaris}}
is a 560-node HPE Apollo 6500 Gen 10+ based system operated by Argonne Leadership Computing Facility (ALCF). There are 40 racks organized in three rows with 16, 12, and 12 racks, respectively. In each rack, there are seven chassis, each containing two nodes. Each node is equipped with a single 2.8 GHz AMD EPYC Milan 7543P 32-core CPU and four NVIDIA A100 GPUs in HGX platform connected via NVLink, two local 1.6TB of SSDs in RAID0 and two Slingshot 11 network adapters. Polaris has \num{2240} GPUs in total.

\vspace{1mm}
\noindent {\bf Perlmutter~\cite{perlmutter}}
is a HPE Cray EX supercomputer with AMD EPYC CPUs and NVIDIA A100 GPUs operated by the National Energy Research Scientific Computing Center (NERSC). Initially, Perlmutter has 14 GPU compute cabinets in total, each segmented into eight chassis. Each chassis contains eight compute blades, each with two nodes. There are \num{1536} NVIDIA A100 GPUs (40G) nodes and 256 NVIDIA A100 GPUs (80G) nodes. 
Each node contains a single AMD EPYC Milan 7763 64-core CPU and four NVIDIA A100 GPUs connected via PCIe, along with four HPE Slingshot 11 NICs. 
More GPUs are added to the cluster and at the time of error log collection, Perlmutter has 7604 GPUs in total. 

\subsection{GPU Architecture}
All three supercomputers are equipped with NVIDIA Ampere GPUs which are released in 2020. 
All of the GPUs in Perlmutter and Polaris are NVIDIA A100, while  Delta has 400 A40 GPUs and 449 A100 GPUs.

\vspace{1mm}
\noindent {\bf NVIDIA A100~\cite{nvidia2020ampere}.}
The NVIDIA Ampere architecture A100 GPU contains seven Graphics Processing Clusters (GPCs), each with seven or eight Texture Processing Clusters (TPCs), and each TPC has two Streaming Multiprocessors (SMs), which is a total of 108 SMs sharing the 40 MB L2 cache. Each SM contains 64 FP32 CUDA Cores and four Tensor Cores, resulting in a total of \num{6912} FP32 CUDA Cores and 432 Tensor Cores per GPU. The A100 GPU includes 40~GB of fast HBM2 DRAM memory on its SXM4-style circuit board, which is organized as five active HBM2 stacks with eight memory dies per stack. 
The A100 HBM2 memory subsystem supports single-error correction double-error detection (SECDED) error-correcting code (ECC) to protect data. Other key memory structures in A100 are also protected by SECDED ECC, including aches and register files.

\vspace{1mm}
\noindent {\bf NVIDIA A40~\cite{nvidia2020ampere}.}
The NVIDIA Ampere architecture A40 GPU contains 84 SMs, each containing 128 CUDA Cores, four Tensor Cores, and one RT Core, which is \num{10752} FP32 CUDA Cores, 432 Tensor Cores, and 84 RT Cores per GPU. The A40 GPU includes 48~GB of GDDR6 DRAM memory with ECC protection.

\subsection{Data Collection Method and Datasets}
 
There are various types of GPU-related errors, to name a few, Dynamic Random Access Memory (DRAM) Single-Bit Errors (SBEs) and Double-Bit Errors (DBEs) reported by Error Correction Codes (ECC), NVLink errors, and Dynamic Page Retirement errors.
Our focus is primarily on memory errors in the DRAM identified and reported by ECC.
We use Data Center GPU Manager (DCGM)~\cite{dcgm}, a suite of tools to manage and monitor GPUs developed by NVIDIA in data centers to monitor SBEs and DBEs in GPU memory.
DCGM regularly updates the aggregated counts of ECC-reported SBEs and DBEs for each GPU, i.e., tracks the historical records of error count.
New errors are identified by observing increases in this error count.
The collected dates and frequency of the error logs studied in this work are summarized in Table~\ref{tab:cluster-info}.

It is essential to distinguish between two terms: {\it error occurrence event} and {\it error count}.
In one error occurrence event where there is an increased error count, multiple (single- or double-bit) errors may be observed in this single error occurrence event.
Distinguishing these two terms is necessary, especially when dealing with DBEs that are detectable but uncorrectable.
When a DBE event happens, upon the detection of DBE(s), ECC would throw an exception and trigger the termination of the GPU application, irrespective of the quantity of DBEs identified.
Therefore, while analyzing the number of detected DBEs can shed light on the severity of such errors, the number of error occurrence events affects the overall cost of addressing DBEs. 
The interarrival time between error occurrence events can be averaged to calculate the Mean-Time-Between-Errors (MTBE), an important metric when accessing system resilience.

\vspace{0.5ex}
\noindent{\textbf{Delta Cluster Error Dataset.}} This dataset has rich information covering a period of more than one year with 849 GPUs, which equates to over 7.32 million GPU device-hours of data. 
Each recorded error occurrence event includes  event type  (i.e., SBE or DBE), timestamp, associated node, and GPU involved in the event.
Note that within Delta, there are 400 NVIDIA A40 GPUs and 449 NVIDIA A100 GPUs, but we do not have the exact mapping of the GPU IDs to their GPU type. We acknowledge this as a limitation of our current study.

\vspace{0.5ex}
\noindent{\textbf{Polaris Cluster Error Dataset.}} 
We record and analyze Polaris errors across 259 days with \num{2240} GPUs in the cluster. 
The missing data from 12/15/2023 to 06/30/2024 is due to lack of access to the database. 
This results in a total of more than 14.66 million GPU device-hours of data. 
On Polaris, only DBEs are recorded. There is no information about SBEs. 

\vspace{0.5ex}
\noindent{\textbf{Perlmutter Cluster Error Dataset.}} 
The Perlmutter dataset contains the GMU memory error information across 326 days with \num{7604} GPUs, resulting in a total of more than 45.79 million GPU device-hours of data.
Several months of data are permanently lost due to database failures.
Of the three studied clusters, Perlmutter is the largest.

\begin{table*}
    \centering

    \caption{Basic characteristics of the three supercomputers studied in this work.}
    \begin{tabular}{|c|c|c|c|}
    \hline

        Cluster & Delta & Polaris & Perlmutter\\ \hline \hline
Node Count & 207 & 560 & 1901 \\ \hline
GPU Count & 849 & 2240 & 7604 \\ \hline  \hline
\multirow{3}{*}{\makecell{Log Collection\\Dates}} & \multirow{3}{*}{12/16/2022 -- 01/07/2024} & \multirow{3}{*}{\makecell{10/01/2023 -- 12/14/2023\\ 07/01/2024– 12/31/2024}} & 07/01/2023 -- 09/30/2023 \\ 
&&& 11/01/2023– 12/20/2023 \\ 
&&&08/01/2024– 01/31/2025 \\ \hline
Log Length & 388 days & 259 days & 326 days \\ \hline
Log Frequency & Every minute & Every 4 seconds & Every hour \\ \hline  \hline
GPU Hours & 7.32 million & 14.66 million & 45.79 million\\

\hline
        
    \end{tabular}
    \label{tab:cluster-info}
\end{table*}

\subsection{Limitations and Scope}
Despite our thorough data collection and analysis efforts, our study is subject to certain limitations.
In this section, we discuss these constraints and clarify the scope of this study.

Firstly, we only have error information on the recorded SBE and DBE errors,  thus this study exclusively focuses on the error occurrences in supercomputers.
Due to privacy concerns, the workload and job scheduling information is not available. Hence, the correlation between hardware failures and software applications cannot be studied in this work.
We partially compensate  this limitation by analyzing the correlation of errors with GPU power consumption, temperature, and utilization, which can reflect the workload patterns.
Limited data availability restricts our capacity to explore further insights into the systems.
SBEs are not recorded on Polaris, which limits the SBE behavior study to  Delta and Perlmutter.

Moreover, due to the limited time of data collection, it is possible that our conclusions might not fully capture the behavior of GPU errors outside this period.
There are 4 days of missing data in Delta, due to possible system failure or maintenance. 
For Polaris, we only have access to the 259 days listed in \autoref{tab:cluster-info}.
For Pulmutter, there are several months of data missing due to database storage error.
Despite the missing data, each cluster still contains several hundred days of logged data, totaling 67.77 million GPU device-hours. Therefore, this study holds statistical significance.

We focus on NVIDIA Ampere GPUs, because they are still one of the main accelerators used in top-100 supercomputers~\cite{top500} and the three supercomputers still have over 50\% utilization.
Although NVIDIA Hopper GPUs are becoming popular in supercomputers, we do not consider them, as they are still in its early stage of deployment where the utilization is low (around 20\%)~\cite{cui2025characterizing}. We do not have access to obtain the error logs of Hopper GPUs.

The primary focus of this work is error behavior characterization. We aim to derive insightful observations and take-away messages to share with the community. The prediction and mitigation of errors fall outside the scope of this study and are subject to future work.

\section{Cross-Supercomputer Error Characterization}
\label{sec:results}

\newcounter{observnum}
\setcounter{observnum}{1}

In this section, we present a detailed characterization study of GPU memory errors across three supercomputers, all equipped with NVIDIA Ampere GPUs. 
We focus on the following aspects:

\begin{enumerate}
    \item {\bf Overall error behavior} (\autoref{sec:results-error-rate}).
    We start with presenting an overview of memory error severity through an analysis of overall error rates and daily error occurrences.
    The observed error rates on different clusters can vary by over three orders of magnitude and we observe burty errors in all three clusters.

    \item {\bf Interarrival Time of Errors} (\autoref{sec:results-mtbe}).
    We quantify the mean-time-between-errors (MTBE) of clusters to assess the frequency of users encountering such errors.
    We analyze the distribution of error interarrival times to provide further insights into the burstiness of errors.
    Our in-depth study reveals that both bursty error patterns and supercomputer scale can affect the characteristics of error interarrival times.

    \item {\bf Correlation with environmental factors} (\autoref{sec:results-correlation}). We analyze the correlation between potential contributing factors including temperature, power, GPU  utilization, and the cooling mechanisms. For most environmental factors,
    we observe weak correlations with errors. No strong correlation is observed. 

    \item {\bf Spatial and temporal analysis} (\autoref{sec:results-spatial}). We analyze the spatial and temporal characteristics of errors  from a supercomputer management and maintenance perspective to identify GPUs that may pose reliability concerns.
    Our analysis reveals that GPU memory errors exhibit spatial and temporal correlation, and the set of erroneous GPUs changes over time.

\end{enumerate}

\subsection{Overview of Errors}
\label{sec:results-error-rate}
The overall numbers of SBEs and DBEs observed in the three clusters during the study period are shown in~\autoref{tab:cluster-errors}. 
The three clusters are presented in ascending order of scale: Delta, Polaris, and Perlmutter, with \num{849}, \num{2240}, and \num{7604} GPUs, respectively.
Comparing the error rates, Perlmutter exhibits a higher SBE rate (5.34$\times$ higher than that of Delta).
The DBE rate of Polaris is notably higher than on the other two clusters: 2555.56$\times$ higher than that of Delta, and 8.41$\times$ higher than that of Perlmutter.
In general, Delta is the most reliable one.
The difference highlights the necessity of a cross-supercomputer study: analyzing one cluster can introduce over-estimation or under-estimation of the error severity.

\vspace{0.4cm}
\hspace{-0.15cm}\fbox{\begin{minipage}{0.95\columnwidth}
		\textbf{Observation \theobservnum}: Observed error rates on different clusters can vary by  over three orders of magnitude. Delta is more reliable than Polaris and Perlmutter supercomputers. 
	\end{minipage}}
\vspace{0.4cm}
\stepcounter{observnum}

\begin{figure*}[hbt]
    \centering
        \begin{minipage}{0.33\textwidth}
                       \footnotesize\centering\includegraphics[width=\columnwidth]{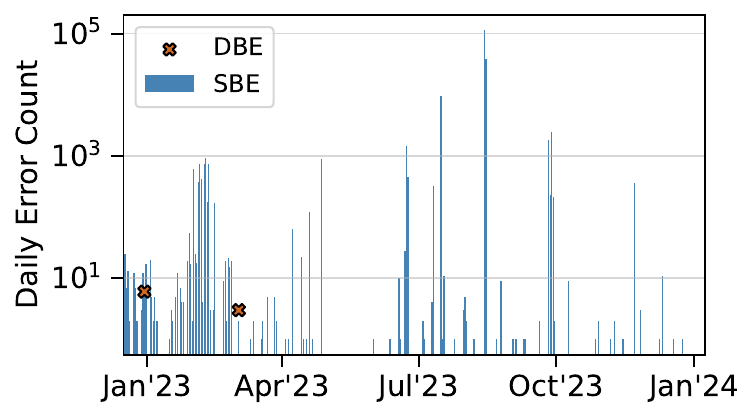}
    (a) Delta (log scale)
        \end{minipage}
        \begin{minipage}{0.33\textwidth}
            \footnotesize\centering\includegraphics[width=\columnwidth]{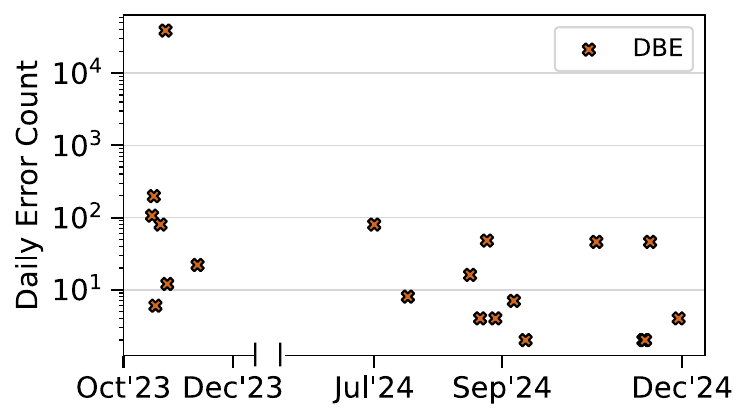}
    (b) Polaris (log scale)
        \end{minipage}
        \begin{minipage}{0.33\textwidth}
                       \footnotesize\centering\includegraphics[width=\columnwidth]{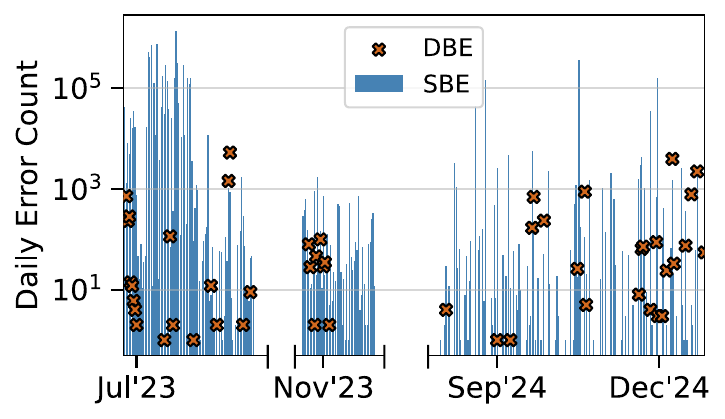}
    (c) Perlmutter (log scale)
        \end{minipage}
    \caption{Daily SBEs and DBEs in the three studied clusters.   Instances of bursty errors are observed in all three clusters.}
    \label{fig:daily-sbe-dbe}
\end{figure*}

In addition to the error rate numbers, we
present the daily error counts for the three clusters in \autoref{fig:daily-sbe-dbe}.
Instances of bursty errors are observed in all three clusters, for example, \num{112656} SBEs on 08/14/2023 in Delta, \num{39144} DBEs on 10/26/2023 in Polaris, \num{1405332} SBEs and 08/06/2023 in Perlmutter. The existence of bursty errors is also observed in another GPU study of the Titan supercomputer~\cite{nie2016large}.

\begin{figure}[tb]
    \centering
    \begin{minipage}{\columnwidth}
		\footnotesize \centering
        
        \footnotesize\includegraphics[width=\columnwidth]{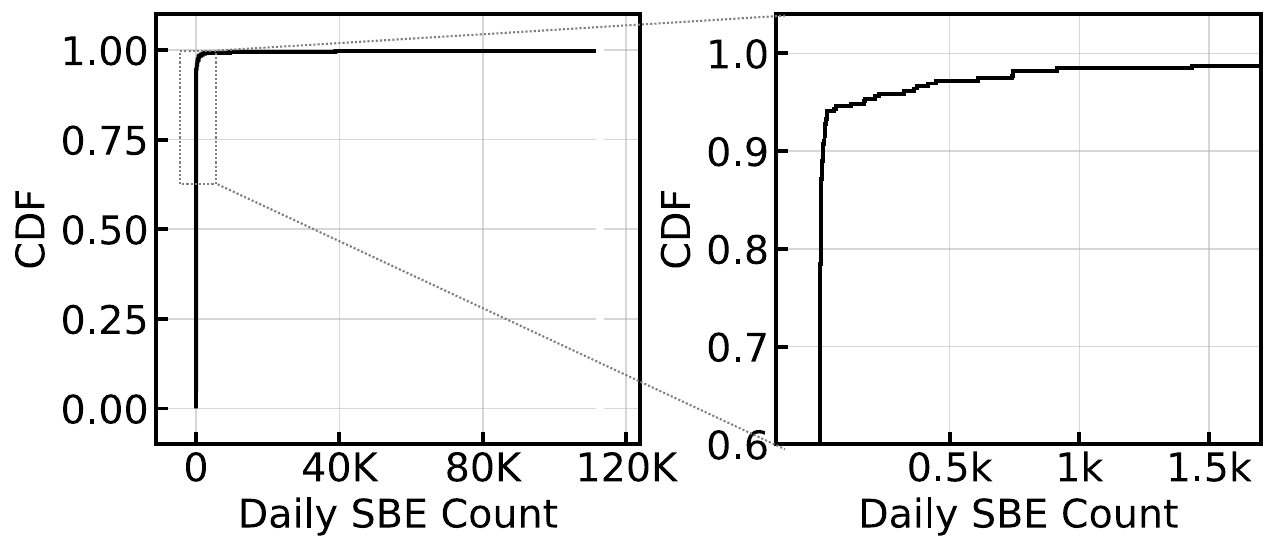}
        
    (a) Delta
    \end{minipage}\hfill
    \centering
    \begin{minipage}{\columnwidth}
		\footnotesize
  \centering\includegraphics[width=\columnwidth]{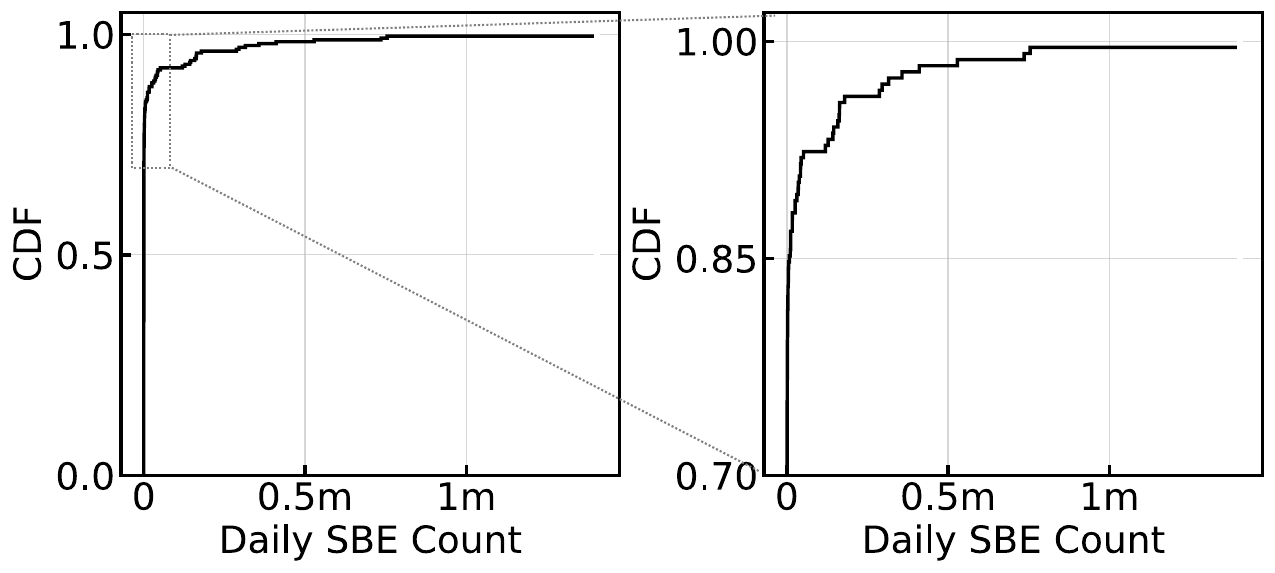}
  (b) Perlmutter
	\end{minipage}

    \caption{CDFs (cumulative distribution functions) of daily SBE count. Both clusters suffer from bursty errors, while Delta exhibits much more severe burstiness.}
    \label{fig:daily-sbe}
    
\end{figure}

For a comprehensive understanding of these bursty error patterns, we calculate and plot the empirical cumulative distribution functions (CDFs) of SBEs in the Delta and Perlmutter  clusters, see \autoref{fig:daily-sbe}.
In Delta,
70.88\%  of days experience no SBEs, indicating that the majority of days are SBE-free.
17.78\% of the monitored days in Delta have less than 10 daily errors.
Additionally, there are outliers: 1.55\% of the days experience more than \num{1000} SBEs,  highlighting the bursty nature of error occurrences.
The highest SBE count is recorded on August 14, 2023, with \num{112656} SBEs.
In Perlmutter,
27.61\% of days are SBE-free, which is much less tha the percentage of SBE-free days in Delta 70.88\%.
The majority, 41.10\% of days, have less than 100 SBEs. 
The number of days exceeding 1000 SBEs in Perlmutter is 19.02\%, which is much higher than Delta. 
The largest daily SBE count in Perlmutter is \num{1405332}.
The high SBE days observed in these clusters confirm the existence of bursty errors.
Moreover, these two distinct daily SBE count distributions emphasize  the importance of our cross-supercomputer perspective to quantitatively understand potential biases in single-supercomputer studies and identify common error characteristics of GPUs.

\begin{figure}[tbh]
    \centering

\begin{minipage}{\columnwidth}
		\footnotesize
  \centering
    
    \includegraphics[width=\columnwidth]{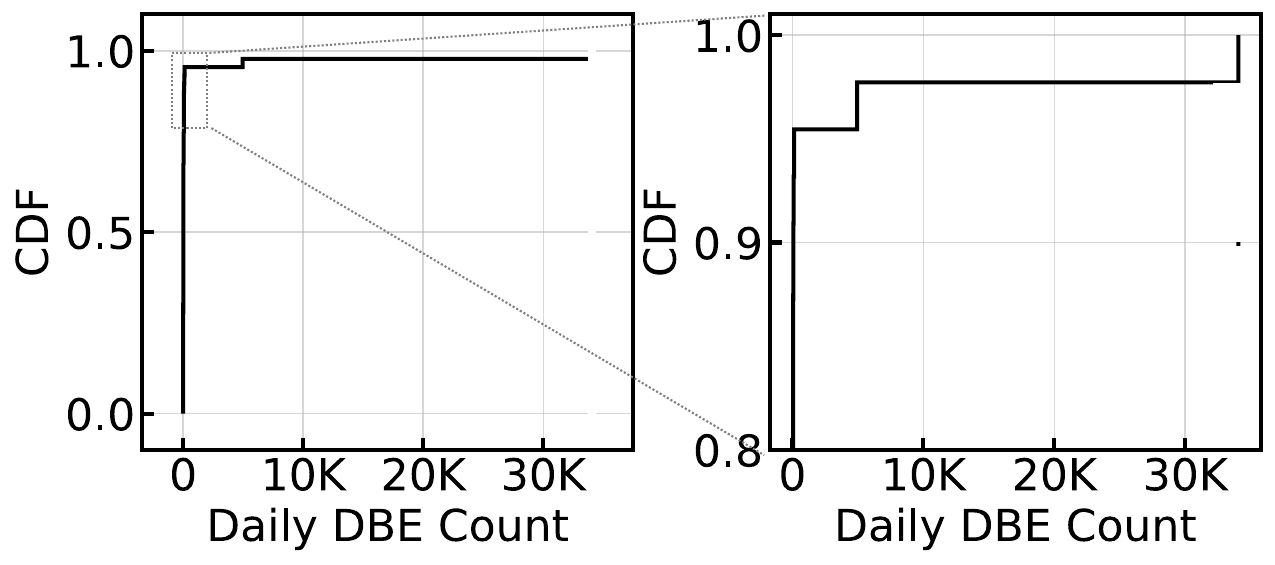}

(a) Polaris
    \end{minipage}\hfill
    \centering
    \begin{minipage}{\columnwidth}
		\footnotesize
  \centering\includegraphics[width=\columnwidth]{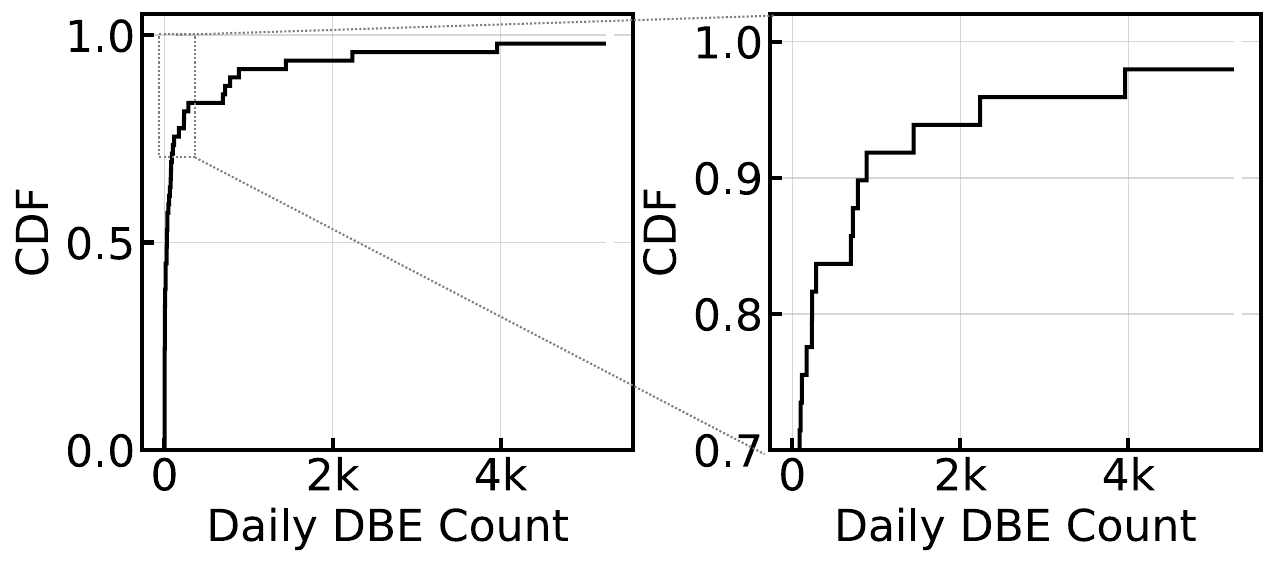}

  (b) Perlmutter
	\end{minipage}

    \caption{CDF of daily DBE count. Burstiness is observed when considering DBEs. 
    }
    \label{fig:daily-dbe}
\end{figure}

Similar bursty error patterns  of DBEs are observed in Polaris and Perlmutter, as shown in \autoref{fig:daily-dbe}.
The DBE-free days are 66.02\% and 84.97\% for Polaris and Perlmutter, respectively.
4.55\% of days in Polaris and 9.32\% in Perlmutter  experience bursty DBEs (more than 1000 errors per day).

\vspace{0.4cm}
\hspace{-0.15cm}\fbox{\begin{minipage}{0.95\columnwidth}
		\textbf{Observation \theobservnum}: Errors are not uniformly distributed over time. Burstiness is generally observed in all three clusters, while their severeness varies.
	\end{minipage}}
\vspace{0.4cm}
\stepcounter{observnum}

Note that while the number of errors and error events are a lot, the actual number of GPUs suffering from errors is not huge, as we show in \autoref{tab:cluster-errors}. Around 4.70\% of the GPUs in both Delta and Perlmutter suffer from SBEs.
In all three clusters, DBE-occuring GPUs are less than 0.53\%.
Recall that DBEs are detected but cannot be corrected by ECC, it is useful to dig out whether SBE occurrence can indicate DBE occurrence.

In Delta cluster, there are only two DBE events observed throughout the year, each on distinct GPUs belonging to different nodes. 
Following its DBE event, one GPU encounters 2 SBEs concurrently.
The other GPU, however, has no SBEs across the whole year. 
Compared to the days with over 1000 SBEs, these two DBE events are not happening on the days that experiencing bursty errors.

\begin{table}[hb]
    \centering   
    \caption{Overview of errors in the studied clusters. }
    
    \begin{tabular}{|c|c|c|c|}
   
    \hline
        {~~~Cluster Name~~~} &  {~~~Delta~~~} &   {~~~Polaris$^*$~~~} & {~~~Perlmutter~~~} \\ \hline \hline
        \# SBEs         & 173936    & N/A       & 7010888  \\ \hline
        \# SBE Events   & 3324      & N/A       & 2016 \\ \hline
        \textbf{SBE Rate}        & \multirow{2}{*}{\textbf{0.53}}  & \multirow{2}{*}{N/A} & \multirow{2}{*}{\textbf{2.83}}  \\ 
        (Per GPU Per Day) & & & \\ \hline \hline
        SBE-occuring GPUs         &  43 (5.06\%)        & N/A       & 344 (4.52\%) \\ \hline \hline
        \# DBEs         &9          & 39837       & 17926 \\ \hline
         \# DBE Events  & 2         & 44         & 77 \\ \hline
        \textbf{DBE Rate}        & \multirow{2}{*}{\textbf{0.000027}}& \multirow{2}{*}{\textbf{0.069}} &  \multirow{2}{*}{\textbf{0.0082}} \\ 
        (Per GPU Per Day) & & & \\ \hline \hline
       DBE-occuring GPUs         &  2 (0.24\%)        & 68 (0.030\%)      & 35 (0.46\%) \\ \hline \hline

       GPUs w/ SBE+DBE         &  1        & N/A      & 29 \\

        \hline
       
    \end{tabular}

    \vspace{2mm}
    { 
    $^*$As SBEs are not recorded in Polaris, those SBE entries are ``N/A''.\hfill
    }
    \label{tab:cluster-errors}
\end{table}

\begin{figure*}[b]
    \centering
    \includegraphics[width=\textwidth]{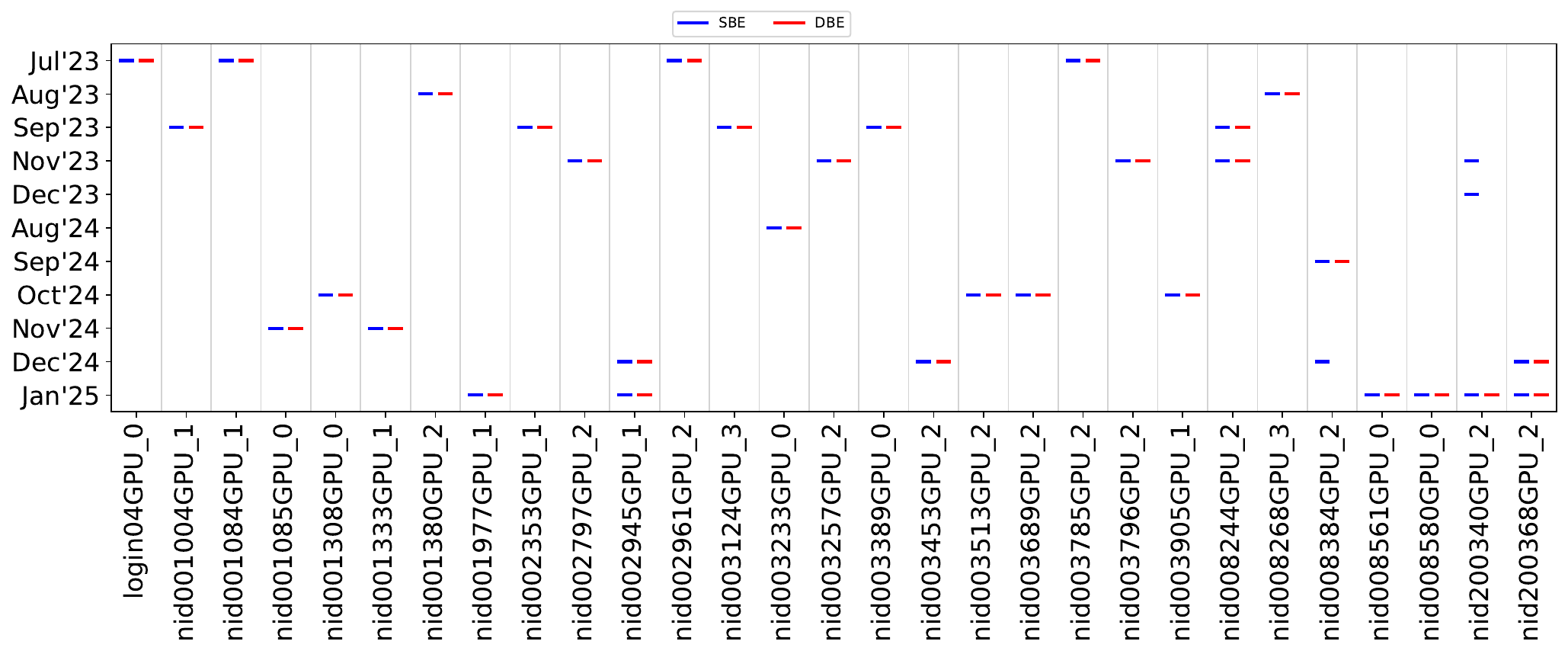}
    \caption{Timeline of GPUs that encounter both SBEs and DBEs in Perlmutter supercomputer. SBEs and DBEs are often correlated.} 
    \label{fig:perlmutter-sbe-dbe-intersection-gpu}
\end{figure*}

For Perlmutter supercomputer, there are 29 GPUs encountering both SBEs and DBEs (see the last row in \autoref{tab:cluster-errors}). 
Their time-wise distribution of SBEs and DBEs are shown in \autoref{fig:perlmutter-sbe-dbe-intersection-gpu}.
We observe that for most of the DBE events, there are SBEs occurring on the same GPU on the same day.
Although the data points are too few to calculate meaningful correlation for each GPU, it is clear that SBEs and DBEs are correlated.
SBEs can be used as an indicator of possible future DBE occurrence. 
This also highlihgts the necessity of studying SBEs despite that it can be protected fully by ECC.

\vspace{0.4cm}
\hspace{-0.15cm}\fbox{\begin{minipage}{0.95\columnwidth}
		\textbf{Observation \theobservnum}: We observe correlation of SBEs and DBEs on the Perlmutter supercomputer. 
	\end{minipage}}
\vspace{0.4cm}
\stepcounter{observnum}

\subsection{Interarrival Time of Errors}
\label{sec:results-mtbe}

We next analyze the error frequency by measuring the interarrival time of errors, i.e., Time Between Errors.
We measure the error interarrival time in the system by calculating the elapsed time between two error-occurring events.
Mean-Time-Between-Errors (MTBE) is then calculated by averaging the interarrival time values and reported in~\autoref{tab:cluster-mtbe}.
To ensure a fair comparison of the interarrival time and MTBE across clusters, we maintain a consistent granularity across all datasets by aggregating error-occurring events into one-hour intervals, matching the log collection frequency of Perlmutter supercomputer (the longest one).
We separately report the MTBE of SBEs and DBEs due to their distinct error rates and consequences, considering that
SBEs are correctable, whereas DBEs would cause program crashes.
We do not calculate the double-bit MTBE for Delta, because there are only two such events.
We also report the standard deviation of MTBE to indicate the distribution of interarrival times.
As shown in \autoref{tab:cluster-mtbe}, Perlmutter exhibits a shorter per-cluster MTBE than Delta (for SBEs) and Polaris (for DBEs), which is consistent with their respective system scales.

\begin{table}[tb]
    \centering   
    \caption{MTBE of the studied clusters.}
    \begin{tabular}{|c|c|c|c|}
 \hline
        \textbf{Cluster Name} &  \textbf{Delta} &   \textbf{Polaris} & \textbf{Perlmutter} \\ \hline
         MTBE (SBE)     & 24.84 $\pm67.85$     & N/A$^*$      & 7.08 $\pm27.31$\\ \hline
        MTBE  (DBE)      &  N/A$^\dagger$               &  $228.23\pm298.27$     &109.97$\pm199.47$\\ \hline

    \end{tabular}

\vspace{3mm}
    {\raggedright
    
    $^*$Polaris logs do not record SBEs.   
    
    $^\dagger$Double-bit MTBE is not computed as Delta logs record only two DBEs.\hfill
    }
    \label{tab:cluster-mtbe}
\end{table}

\vspace{0.4cm}
\hspace{-0.15cm}\fbox{\begin{minipage}{0.95\columnwidth}
		\textbf{Observation \theobservnum}: MTBE is affected by the cluster scale. As the largest cluster, Perlmutter has the lowest MTBE. 
	\end{minipage}}
\vspace{0.4cm}
\stepcounter{observnum}

\begin{figure}[tb]
     \begin{minipage}{\columnwidth}
                       \footnotesize\centering
    \includegraphics[width=\columnwidth]{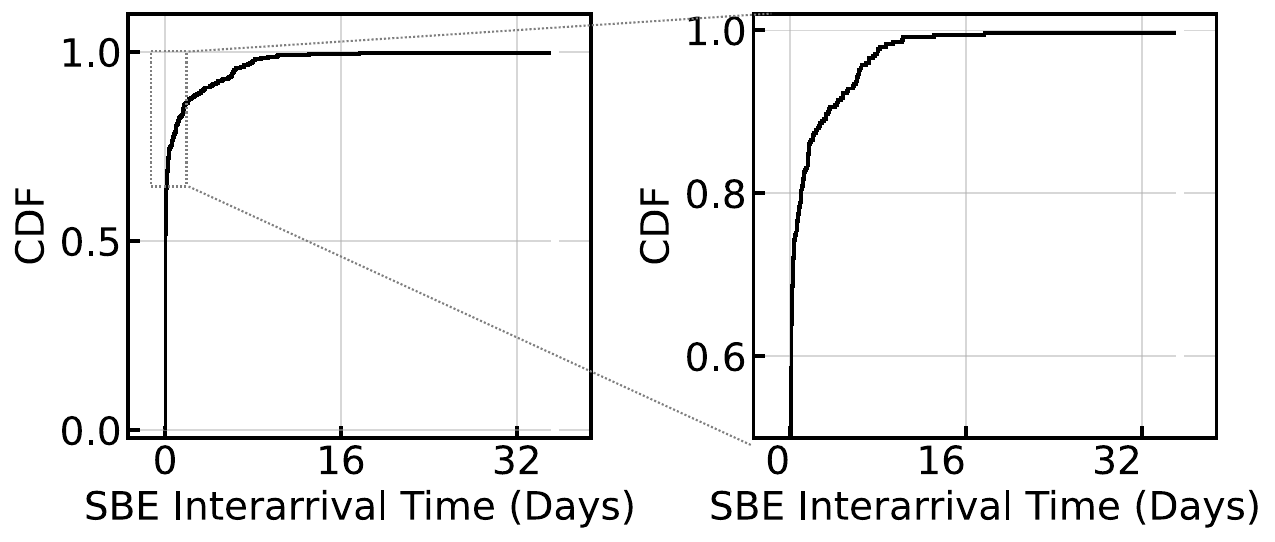}

    (a) Delta
    \end{minipage}\hfill
  \begin{minipage}{\columnwidth}
                       \footnotesize\centering
     \includegraphics[width=\columnwidth]{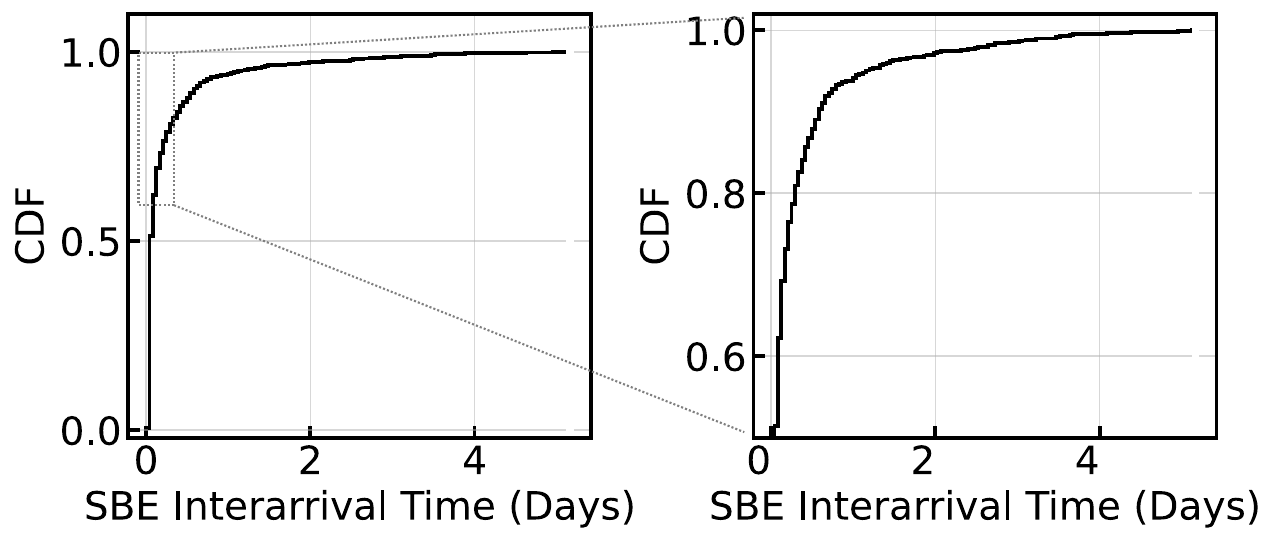}

     (b) Perlmutter
\end{minipage}
    \caption{CDF of SBE interarrival times. Bursty errors are observed and Delta experience more severe burstiness than Perlmutter.}
    
    \label{fig:sbe-interarrival-time-cdf}

\end{figure}

Similar to the overall error rates, one single number of MTBE is not sufficient to represent the overall resilience of a cluster. 
The bursty error patterns are reflected in the large standard deviation values in~\autoref{tab:cluster-mtbe}. 
We investigate the distributions of interarrival times in clusters, as we show in \autoref{fig:sbe-interarrival-time-cdf}.

\autoref{fig:sbe-interarrival-time-cdf}(a) shows the CDF of SBE interarrival times in Delta cluster.
94.31\% of the interarrival times are less than one hour, indicating the bursty error patterns.
Meanwhile, we observe a long tail in the distribution: 0.81\% of the SBE interarrival times exceed 120 hours, extending up to  851 hours (35 days).
This long tail further underscores the bursty error patterns in Delta.

The SBE interarrival times of Perlmutter are presented in \autoref{fig:sbe-interarrival-time-cdf}(b).
Among the SBE interarrival times in Perlmutter, 51.42\% fall below one hour, which is significantly less than the Delta case where the majority (94.31\%) are below one hour, indicating that Delta exhibits more severe bursty patterns than Perlmutter.
The maximum SBE interarrival time in Perlmutter is around 5 days, which is much shorter compared to the one observed in Delta (35 days). 
This disparity is related to the scales of the two supercomputers. 
For a simplified illustration example, Perlmutter is about 8.96$\times$ larger than Delta; even assuming the same GPU error rate, this scale difference implies that encountering an SBE in Perlmutter is approximately 8.96$\times$ more likely than in Delta. 
Therefore, the observed difference of the longest interarrival time  is reasonable and aligns with the expectations set by their scales.
Moreover, these numbers do not reveal a linear relationship between the scale of the cluster and the error interarrival times, likely due to the complex nature of the cluster environment.

\begin{figure}[tb]
        \begin{minipage}{.7\columnwidth}
                       \footnotesize\centering\includegraphics[width=0.9\columnwidth]{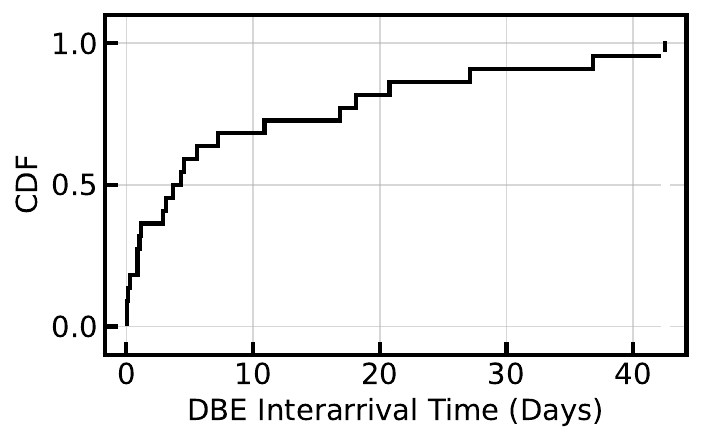}
                       
    (a) Polaris
        \end{minipage}
   \begin{minipage}{.7\columnwidth}
                       \footnotesize\centering\includegraphics[width=0.9\columnwidth]{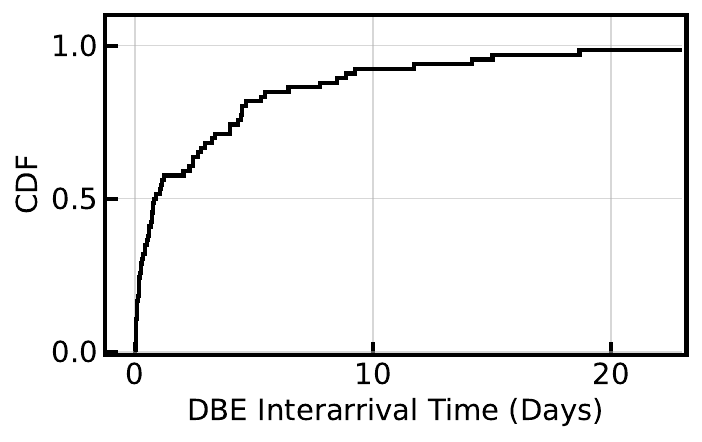}
                       
    (b) Perlmutter
        \end{minipage}

    \caption{CDF of interarrival times of DBEs. Cluster scale affects error characteristics.}

    \label{fig:dbe-interarrival-time}
\end{figure}

\autoref{fig:dbe-interarrival-time} shows the CDF of interarrival times of DBEs.
Given that the DBE rate is significantly lower than the SBE rate, the CDFs of DBE interarrival times are more sparse than the SBE ones and their characteristics are distinct.
In Polaris, 27.27\% of DBE interarrival times are less than one day, while in Perlmutter, the proportion is 51.56\%.
For the distribution tail which indicates long interarrival times, the longest DBE interarrival time in Polaris is 42 days, while the longest one in Perlmutter is 23 days.
All these characteristics are intrinsically linked to the scale and the level of error burstiness of the two clusters.

\vspace{0.4cm}
\hspace{-0.15cm}\fbox{\begin{minipage}{0.95\columnwidth}
		\textbf{Observation \theobservnum}: Both bursty error patterns and supercomputer scale can affect the characteristics of error interarrival times. Relying on a single number, MTBE, may introduce bias into resilience estimation. It is necessary to consider the distribution of error interarrival times to obtain an accurate and holistic resilience assessment. 
	\end{minipage}}
\vspace{0.4cm}
\stepcounter{observnum}

\begin{figure*}[htb]
    \centering
      \begin{minipage}{.75\columnwidth}
    \footnotesize\centering\includegraphics[width=\columnwidth]{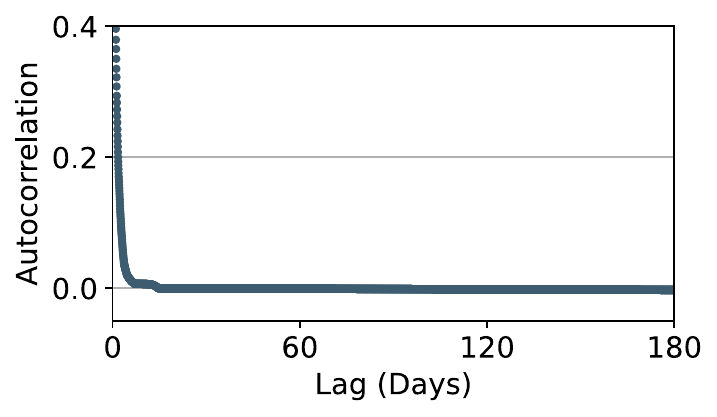}
    (a) Delta, SBE   
        \end{minipage}
        \begin{minipage}{.75\columnwidth}
            \footnotesize\centering\includegraphics[width=\columnwidth]{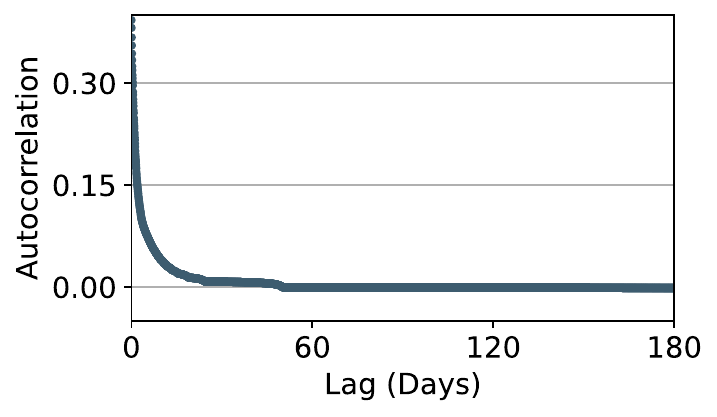}
    (b) Perlmutter, SBE
        \end{minipage}

      \begin{minipage}{.75\columnwidth}
    \footnotesize\centering\includegraphics[width=\columnwidth]{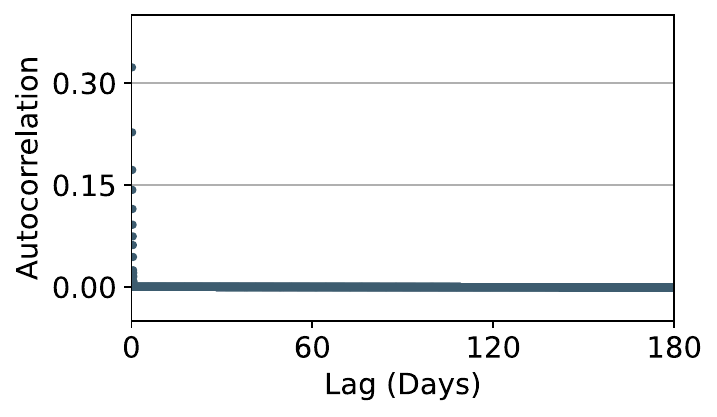}
                       
    (c) Polaris, DBE
        \end{minipage}
        \begin{minipage}{.75\columnwidth}
            \footnotesize\centering\includegraphics[width=\columnwidth]{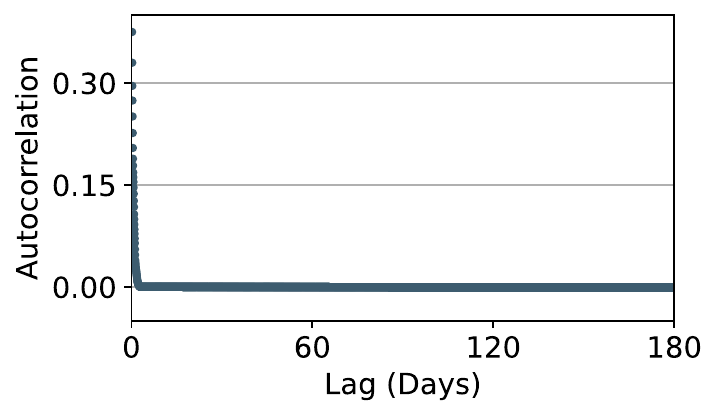}
    (d) Perlmutter, DBE
        \end{minipage}
    \caption{Autocorrelation of the DBE interarrival times. The high autocorrelation within a short lag indicates the bursty error patterns. There is no periodicity of errors observed.} 
    \label{fig:delta-autocorrelation}
\end{figure*}

We explore the potential temporal correlation and periodicity of errors leveraging the autocorrelation function of error interarrival times.
Autocorrelation evaluates the similarity between a time series and its lagged version and is always in the range of $[-1,1]$~\cite{leemis2006discrete}.
A higher positive number refers to a stronger correlation, suggesting a higher level of periodicity with a certain period length, i.e., the lag.
Zero values indicate the absence of correlation at the given lag and negative autocorrelation numbers point to the opposite relationship with its lagged series.
We vary lags by hour to compute the autocorrelation values and explore possible periodicity.
\autoref{fig:delta-autocorrelation} shows the autocorrelation of error interarrival times.
High autocorrelations are observed with lags of less than 24 hours, which confirms the large body of bursty errors in all clusters.
With lags increasing, we do not observe any notable correlation. 
We conclude that burstiness is severe in all clusters but no periodicity is observed.

\vspace{0.4cm}
\hspace{-0.15cm}\fbox{\begin{minipage}{0.95\columnwidth}
		\textbf{Observation \theobservnum}: No periodicity of errors is observed in clusters.
	\end{minipage}}
\vspace{0.4cm}
\stepcounter{observnum}

\begin{table}
    \centering
    \caption{Correlation of errors and environmental factors. No strong correlation is observed.
}

\centering

    \begin{tabular}{|c|c|c|c|c|}
    \hline
         \multicolumn{2}{|c|}{} & Temperature & Power & GPU Utilization     \\ \hline
         
         \multirow{1}{*}{Delta} & SBE & 0.14 & 0.14 & 0.20    \\
          \hline

         \multirow{1}{*}{Polaris} 
          & DBE & 0.089 & 0.17 & 0.18    \\ \hline

         \multirow{2}{*}{Perlmutter} & SBE & 0.011 & 0.23 & N/A$^*$     \\
          & DBE & 0.19 & 0.32 & N/A$^*$    \\ \hline

    \end{tabular}
    
    \vspace{3mm}
    {
    $^*$ Perlmutter logs do not record GPU utilization data. \hfill
    }
    \label{tab:temp-util-power-corr}
\end{table}

\subsection{Correlation with Environmental Factors}
\label{sec:results-correlation}

In this section, we examine the correlation of SBE/DBE occurrences with environmental factors that may be related to error behavior.
Specifically, we consider  temperature, power consumption, and GPU utilization.
Additionally, we also discuss the liquid cooling systems used in the three supercomputers and their impact on GPU error behavior.

\autoref{tab:temp-util-power-corr} summarizes  the correlation of SBEs and DBEs with these factors across the three clusters. 
While we do not have access to the detailed workload information (due to privacy issues), these environmental factors can serve as indirect indicators of workload patterns.  
Overall, we do not observe strong correlations.
While weak -- but not entirely negligible -- correlations are present in most of the  cases, 
except that the DBEs in Polaris and the SBEs in Perlmutter show no observable correlation with temperature.
It is inherently complex to design and build a supercomputer, thus the sources and contributing factors of memory errors are also complicated,
including, chip manufacturing variability, physical node placement, and other hardware-level influences.
As such, we are not able to draw definitive conclusions regarding the influence of environmental factors.
Nonetheless, we suggest prioritizing GPUs  that consistently exhibit high temperature, power consumption, and GPU utilization for proactive reliability  management.

\begin{figure*}[tb]
    \centering
    \includegraphics[width=0.7\textwidth]{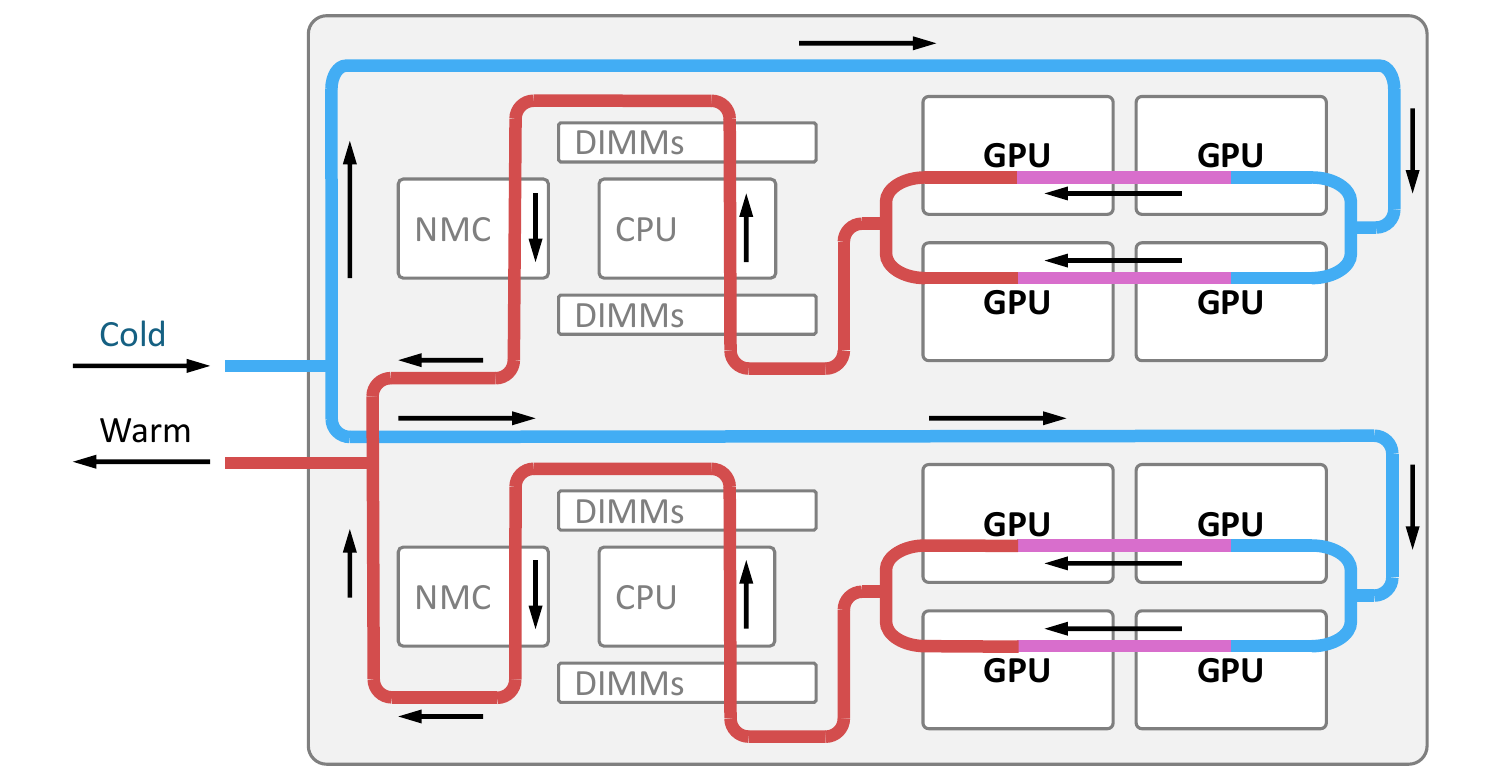}
    \caption{The liquid cooling flow of one Perlmutter blade. Each blade contains two nodes. Two GPUs share one cooling loop. The other two systems, Delta and Polaris, have a similar direct liquid cooling system. } 
    \label{fig:cooling}
\end{figure*}

 \begin{table}[b]
    \centering   
    \caption{Average temperature in 3 clusters.}
    \begin{tabular}{|c|c|c|c|}
   
    \hline
         GPU ID & Delta & Polaris & Perlmutter \\ \hline\hline
         0 & \multirow{4}{4em}{\centering N/A$^\dagger$} & 36.43 & 29.56    \\ \cline{1-1} \cline{3-4}
         1 &  & 39.88 & 28.48      \\ \cline{1-1} \cline{3-4}     
         2 &  & 36.38 & 29.57      \\ \cline{1-1} \cline{3-4}
         3 &  & 41.18 & 28.49      \\ \hline\hline

        Average & 35.66 & 38.47& 29.03 \\ \hline
         
    \end{tabular}

    \vspace{3mm}
    {
     $^\dagger$ We do not have information of the GPU ID mapping in Delta.
    
    }
    \label{tab:gpu-temperature}
\end{table}

All three supercomputers leverage  direct liquid cooling systems with similar configurations.  
For brevity, here we present the liquid cooling flow of a single Perlmutter blade as an example, see \autoref{fig:cooling}.
Each blade hosts two nodes, each with one CPU and four GPUs.
Within each node, two GPUs share one cooling loop.
Additionally, we calculate the average temperature of GPUs, shown in \autoref{tab:gpu-temperature}.
On average, Perlmutter shows the lowest temperatures, followed by Delta, while Polaris exhibits the highest.
On both Polaris and Perlmutter,  GPUs 0 and 1 share one cooling loop and GPUs 2 and 3 share the other.
For Polaris, the loop starts with the lower-numbered GPU, whereas in Perlmutter, the loop begins with the higher-numbered GPU.
This is confirmed by the temperature values in \autoref{tab:gpu-temperature}.
Recall the error rate statistics in \autoref{tab:cluster-errors}, Delta is the most reliable system, although here the average temperature of Delta is much higher than Perlmutter.
Still, we cannot draw any conclusive relationship between  temperature, cooling system design, and error characteristics.

\vspace{0.4cm}
\hspace{-0.15cm}\fbox{\begin{minipage}{0.95\columnwidth}
		\textbf{Observation \theobservnum}: For most environmental factors,
    we observe weak correlations with errors. No strong correlation is observed.
	\end{minipage}}
\vspace{0.4cm}
\stepcounter{observnum}

\subsection{Spatial and Temporal Behaviour}
\label{sec:results-spatial}

From a supercomputer management and maintenance perspective, the spatial and temporal reliability characteristics of different GPUs are of interest to cluster administrators.
We start with characterizing the transition of spatial error behavior  across time in clusters, then we further investigate the severity of GPU errors over time.

\begin{figure}[b]
    \centering
    \vspace{-\baselineskip}
    
    \includegraphics[width=0.7\columnwidth]{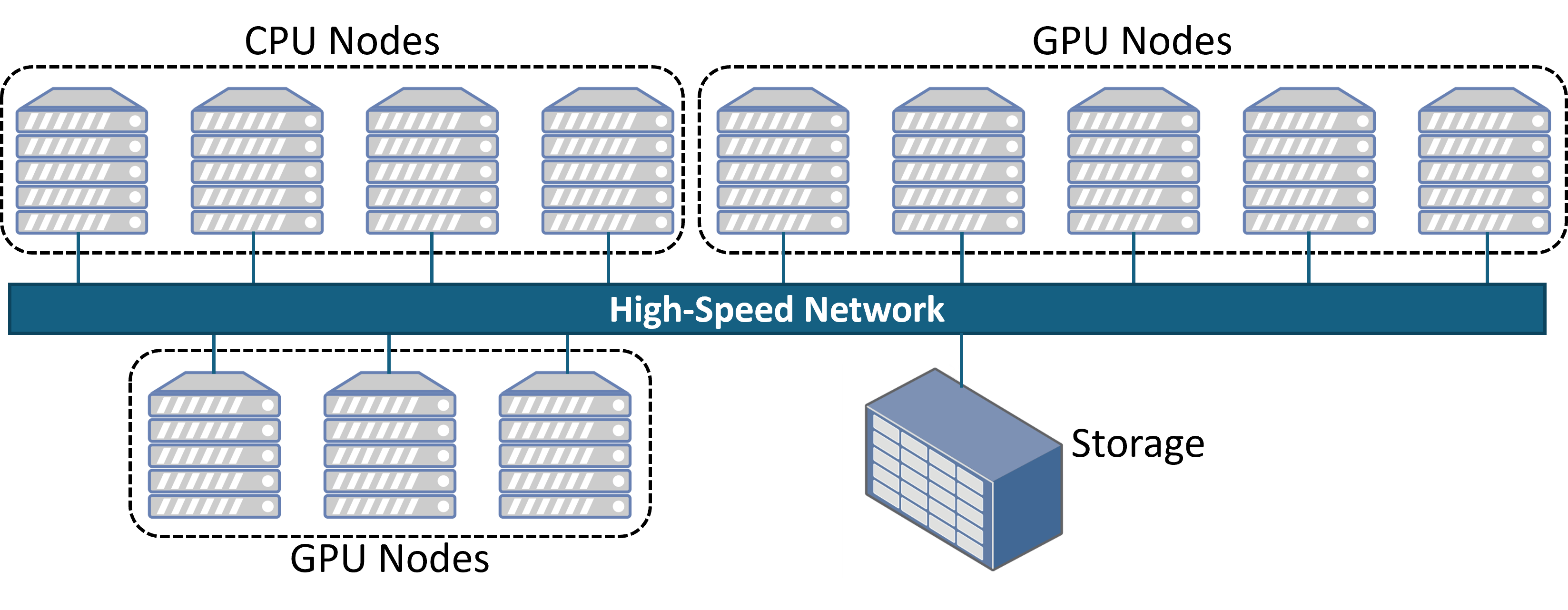} 
    \caption{Physical location of nodes in Delta. }
    \label{fig:delta-arch}
    \vspace{-\baselineskip}

    \end{figure}

\begin{figure}[h]
\vspace{\baselineskip}
    \centering
              \begin{minipage}{\columnwidth}
                       \footnotesize\centering\includegraphics[width=0.23\columnwidth]{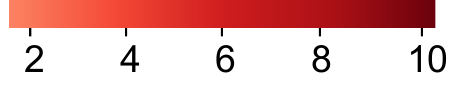}
    \vspace{1mm}
        \end{minipage}
        
            \begin{minipage}{0.23\columnwidth}                       \footnotesize\centering\includegraphics[width=0.7\columnwidth]{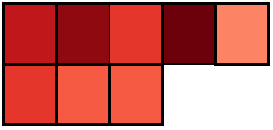}
    
    (a) Monthly Avg.
        \end{minipage}
                    \begin{minipage}{0.23\columnwidth}
                       \footnotesize\centering\includegraphics[width=0.7\columnwidth]{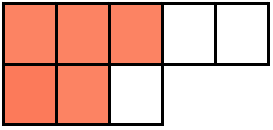}
    
    (b) January
    
        \end{minipage}            
        \begin{minipage}{0.23\columnwidth}
                       \footnotesize\centering\includegraphics[width=0.7\columnwidth]{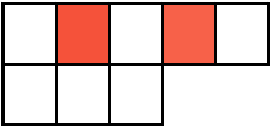}
    
    (c) August
    
        \end{minipage}
                    \begin{minipage}{0.23\columnwidth}
                       \footnotesize\centering\includegraphics[width=0.7\columnwidth]{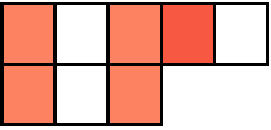}
    
    (d) September    
        \end{minipage}

    \caption{Number of error-occurring days within each cabinet over time. Hotspots vary over time.}
    \label{fig:delta-error-occuring-days}
    
\vspace{\baselineskip}

    \centering
              \begin{minipage}{\columnwidth}
                       \footnotesize\centering\includegraphics[width=0.23\columnwidth]{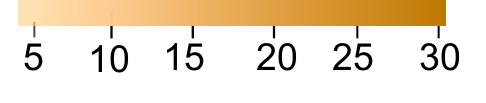}
    \vspace{1mm}
        \end{minipage}
        
            \begin{minipage}{0.23\columnwidth}                       \footnotesize\centering\includegraphics[width=0.7\columnwidth]{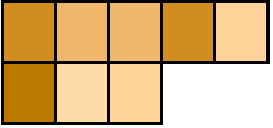}
    
    (a) Monthly Avg.
        \end{minipage}
                    \begin{minipage}{0.23\columnwidth}
                       \footnotesize\centering\includegraphics[width=0.7\columnwidth]{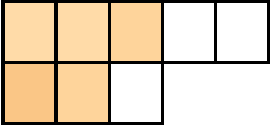}
    
    (b) January
    
        \end{minipage}            
        \begin{minipage}{0.23\columnwidth}
                       \footnotesize\centering\includegraphics[width=0.7\columnwidth]{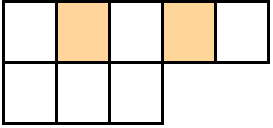}
    
    (c) August
    
        \end{minipage}
                    \begin{minipage}{0.23\columnwidth}
                       \footnotesize\centering\includegraphics[width=0.7\columnwidth]{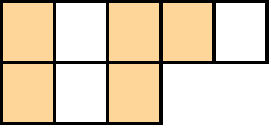}
    
    (d) September    
        \end{minipage}

    \caption{Number of SBE-occurring GPUs within each cabinet over time. Although the heatmap pattern changes slightly  comparing to the characteristics when considering the number of error-occurring days shown in \autoref{fig:delta-error-occuring-days}, the same observation preserves: hotspots vary over time.}
    
    \label{fig:delta-error-occuring-gpu}
\end{figure}

We first investigate the spatial correlation of errors: in particular, whether GPUs within the same cabinet exhibit similar error characteristics.
At the time of writing this paper, we only have the full machine topology for Delta, shown in \autoref{fig:delta-arch}.
In Delta, 207 GPU nodes with a total of 849 GPUs are organized in eight cabinets.
We leverage this architectural information to determine the
spatial distribution of SBEs in Delta.
Given that bursty errors are commonly observed in Delta, instead of presenting the number of SBEs in each cabinet, we calculate the number of days that a cabinet encounters errors (\autoref{fig:delta-error-occuring-days}) and the number of SBE-occurring GPUs within a cabinet (\autoref{fig:delta-error-occuring-gpu}).
For brevity, we select three time windows (in February, June, and August) to present the dynamics of the physical distribution of errors.
We observe that hotspots of SBEs vary across different months.
This trend is consistent considering the two metrics, error-occurring days and error-occurring GPUs. 
This variation indicates physical correlations among errors, with the SBE-occurring cabinets changing over time.

\vspace{0.4cm}
\hspace{-0.15cm}\fbox{\begin{minipage}{0.95\columnwidth}
		\textbf{Observation \theobservnum}: GPU memory errors are spatially and temporally correlated in the Delta supercomputer.
	\end{minipage}}
\vspace{0.4cm}
\stepcounter{observnum}

\begin{figure}[htb]
    \centering
    \includegraphics[width=0.65\columnwidth]{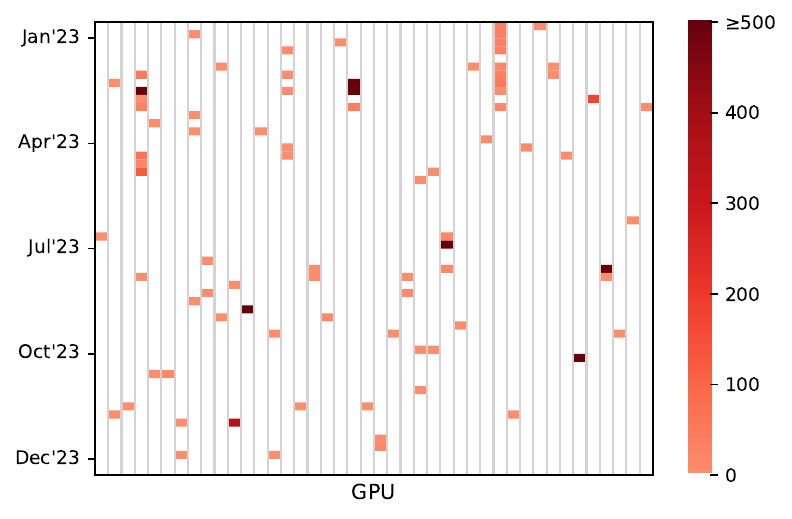} 
    \caption{Number of SBEs observed on SBE-occuring GPUs over time in Delta. Most of the GPUs encounter SBEs for only 1--3 weeks. There is no GPU that continually encounters SBEs throughout the whole year.}
    \label{fig:delta-heatmap}
\end{figure}

\begin{figure*}[tb]
    \centering
    \begin{minipage}{\textwidth}
		\footnotesize
  \centering\includegraphics[width=\textwidth]{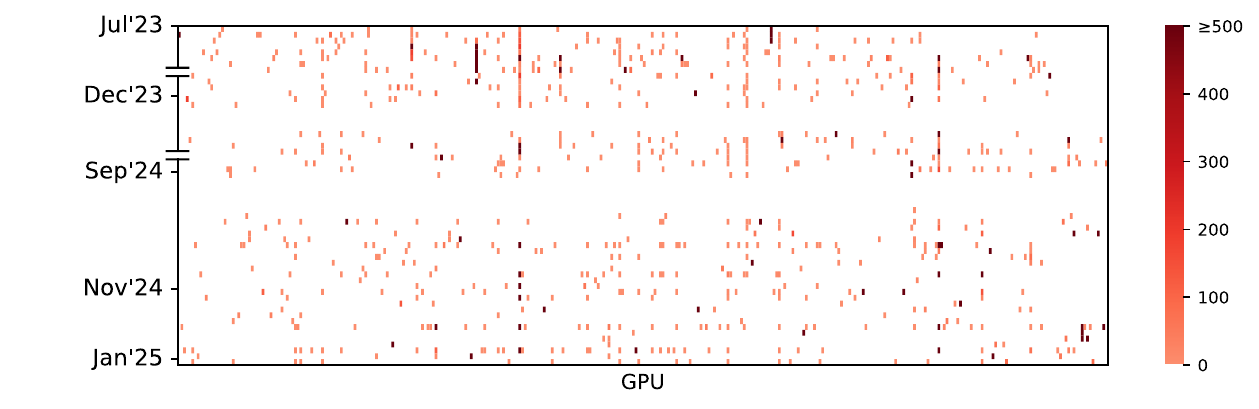}
	\end{minipage}
 	
   \vspace{-\baselineskip}
    \caption{Per-GPU SBE counts over time in Perlmutter. Here we  show all the SBE-occurring GPUs. Erroneous GPUs vary over time.}
   \vspace{-\baselineskip}

    \label{fig:perlmuter-sbe-heatmap}
\end{figure*}

\begin{figure}[tb]
    \centering
\begin{minipage}{0.69\columnwidth}
  \footnotesize\centering\includegraphics[width=0.95\columnwidth]{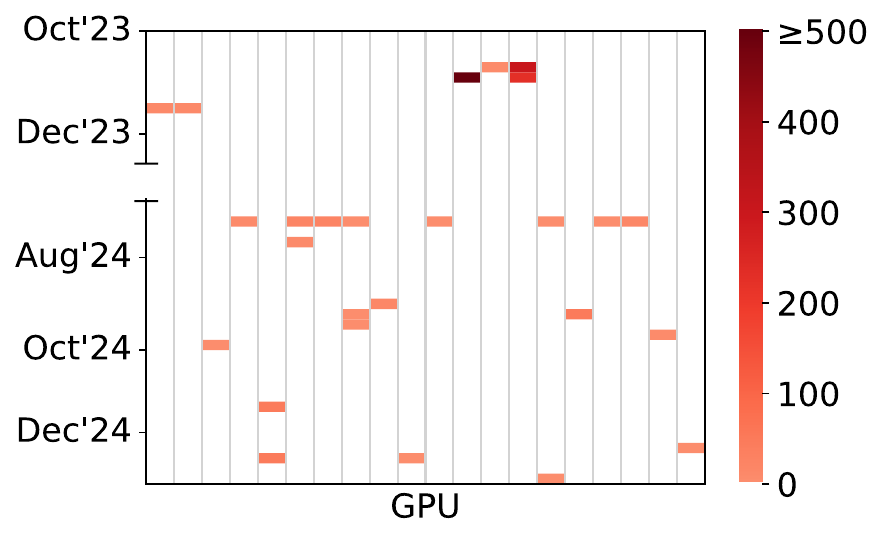}

    (a) Polaris
	\end{minipage}\hfill
    \begin{minipage}{0.69\columnwidth}
  \footnotesize\centering\includegraphics[width=\columnwidth]{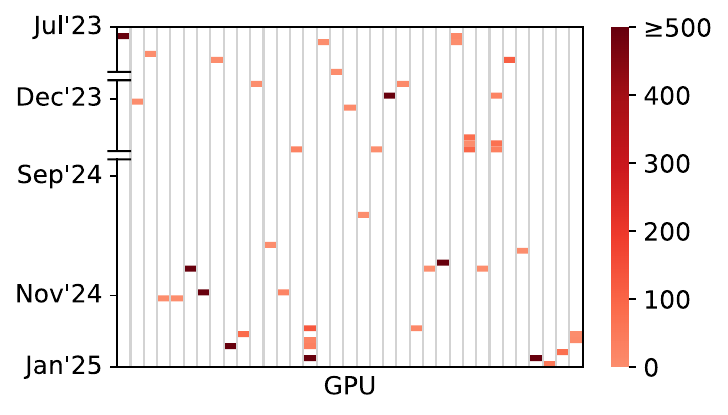}

    (b) Perlmutter
	\end{minipage}
 \caption{Per-GPU DBE counts observed over time on Polaris and Perlmutter. There are less DBE-occurring GPUs than the case of SBEs. The same observation can be derived: DBE-encountering GPUs vary over time.
 }
    \vspace{-\baselineskip}
 
    \label{fig:dbe-heatmap}
\end{figure}

We further investigate the severity of GPU errors over time.
\autoref{fig:delta-heatmap} shows the number of SBEs over time observed in SBE-occurring GPUs in Delta. For ease of reading, we present the weekly SBE count, with darker colors indicating more errors observed in a GPU (x-axis) during a certain week (y-axis). 
Most of the GPUs suffer from SBEs for only 1--3 weeks.
We do not observe any GPU that continually encounters SBEs throughout the whole year.
Similar observations are drawn from the other SBE and DBE datasets, as depicted in \autoref{fig:perlmuter-sbe-heatmap} (SBEs in Perlmutter) and \autoref{fig:dbe-heatmap} (DBEs in Polaris and Perlmutter clusters).
In general, GPUs with high error counts tend to vary over time, suggesting that
for system monitoring and maintenance, focusing on GPUs with historically high error counts may not be a good strategy.

\vspace{0.4cm}
\hspace{-0.15cm}\fbox{\begin{minipage}{0.95\columnwidth}
		\textbf{Observation \theobservnum}: Erroneous GPUs vary over time. For the management of supercomputers, focusing on  GPUs with historically high SBE counts may not be a good strategy. 
	\end{minipage}}
\vspace{0.4cm}
\stepcounter{observnum}

\section{Discussions and Lessons Learned}
~\label{sec:implications}

In this section, we  discuss the opportunities for efficient machine health management and large-scale application checkpointing in HPC systems. 
These opportunities are enabled by our GPU memory error characterization study presented in \autoref{sec:results}.
We also compare the Ampere GPU error statistics with previous generations to observe the trend of GPU reliability.
Specifically, we present quantitative comparisons to the Summit supercomputer with NVIDIA V100 GPUs.

\subsection{Compared to Previous GPU Generations}
~\label{sec:compare-other-gpus}

Large-scale GPU memory error log analyses for supercomputers mainly focus on three systems: 
Summit, with V100 GPUs (the predecessor of A100)~\cite{oles2024understanding},
Titan, with K20X GPUs~\cite{tiwari2015understanding,nie2016large,nie2017characterizing}, and Blue Waters, with K20X GPUs~\cite{di2014lessons}.
We briefly compare observations derived from our A100 GPU datasets with those of previous GPU generations, to seek similarities and differences.
First, the bursty patterns are confirmed in the Titan supercomputer~\cite{nie2016large}.
Second, notable periodicity is observed in Titan~\cite{nie2016large}, however, we do not observe any periodicity in Delta or Perlmutter.

Furthermore, we quantitatively compare the characteristics of DBEs\footnote{The Summit public dataset~\cite{oles2024understanding} of V100 GPU errors does not contain SBE information.} between V100 (in the Summit supercomputer~\cite{oles2024understanding}) and A100 GPUs.
Both generations of GPUs are equipped with HBM2 (High Bandwidth Memory 2).
Note that consecutive DBEs are grouped and considered as one single DBE event in~\cite{oles2024understanding}.
Here we do not cluster consecutive errors because  grouping consecutive errors together filters out the lasting time of these  errors which are meaningful to show the level of severeness of the error consequences.
By processing these datasets through the same way and the same granularity (per hour), we still ensure a fair and meaningful comparison.

\begin{figure}[tb]
    \centering
    \includegraphics[width=0.75\columnwidth]{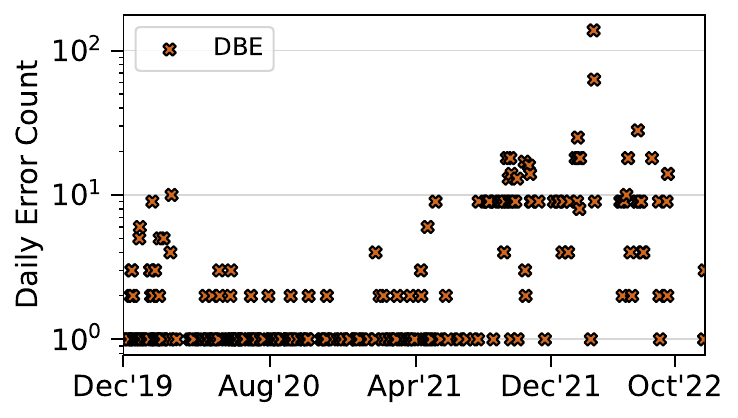} 

   \vspace{-\baselineskip}
    
    \caption{Daily DBE count  (log scale) in the Summit supercomputer, equipped with NVIDIA V100 GPUs. Bursty patterns are observed, but not as severe as the A100 GPUs.}

   \vspace{-\baselineskip}
    
    \label{fig:summit-dbe}
\end{figure}

\autoref{fig:summit-dbe} shows the daily DBE count in Summit, confirming a moderate level of error burstiness. Comparing to the daily DBE count of the three A100 supercomputers (\autoref{fig:daily-dbe}) where we sometimes observe more than 1000 errors per day, the maximum daily error count in Summit is 138.
This indicates that the Summit supercomputer experiences bursty errors, but not as much as the supercomputers accelerated by A100 GPUs.

Furthermore, we calculate the DBE rate and MTBE values and present the statistics in \autoref{tab:summit-dbe-comparison}.
The DBE rate of Summit is higher than Delta but lower than that of Polaris and Perlmutter.
There are more DBE-occurring GPUs in Summit, both in terms of raw numbers and the normalized percentage.
The MTBE of Summit is slightly longer than Perlmutter but lower than Polaris.
The standard deviation of Summit MTBE is much smaller than the A100 clusters, confirming that Summit suffers less from bursty errors. 
In general, we cannot clearly conclude the reliability ranking of these clusters. Given that all four clusters use HBM2 memory, it is reasonable that they share similar memory error characteristics.

\begin{table}
    \centering   
    \caption{Error statistics in the Summit Supercomputer.}
   \vspace{-\baselineskip}
    
    \begin{tabular}{|c|c|}
   
    \hline
        \textbf{~~~Cluster Name~~~} &  \textbf{~~~Summit~~~} \\ \hline \hline

        \# GPUs & 28471 \\ \hline
        \# Logged Days  & 1026 \\ \hline \hline
        \# DBEs         &1088          \\ \hline
         \# DBE Events  & 1008       \\ \hline
        DBE Rate (Per GPU Per Day)       & 0.00022 \\ \hline 

       DBE-occuring GPUs         &  124 (2.6\%)     \\ \hline 

       MTBE & 122.62 $\pm$81.64 \\ \hline
       
    \end{tabular}
   \vspace{-1\baselineskip}

    \label{tab:summit-dbe-comparison}
\end{table}

\vspace{0.4cm}
\hspace{-0.15cm}\fbox{\begin{minipage}{0.95\columnwidth}
		\textbf{Take-away Message 1}:  NVIDIA V100  and A100 GPUs both use HBM2 and share similar memory error characteristics.
	\end{minipage}}
\vspace{0.4cm}

\subsection{Reliable Operation of Supercomputers}

We discuss the insights derived from our characterization study from two major observations. Firstly, the observation of bursty  error patterns suggests GPU ECC monitoring frequency. Secondly, our spatial and temporal analysis provides insights for cluster management and maintenance.

\subsubsection{GPU ECC Monitoring Frequency}
Among the three clusters we studied, GPU ECC errors are monitored at different frequencies.
Polaris error logs are recorded every four seconds, while Delta errors are recorded every minute, and we retrieve Perlmutter errors using a frequency of every hour.
This variation motivates a discussion of the trade-off between the overhead of error monitoring and the ability to promptly identify and react to errors.
Ideally, monitoring should be performed as infrequently as possible while still capturing sufficient information to detect abnormal GPU behaviors, especially bursty errors. 

\begin{figure}[tb]
    \centering
    \begin{minipage}{0.49\columnwidth}
		\footnotesize
  \centering\includegraphics[width=\columnwidth]{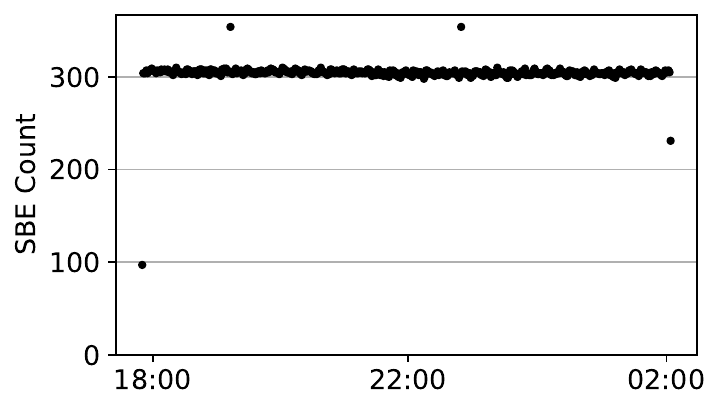}
  
    (a) Per Minute
	\end{minipage}
 	\begin{minipage}{0.49\columnwidth}
  \footnotesize\centering\includegraphics[width=\columnwidth]{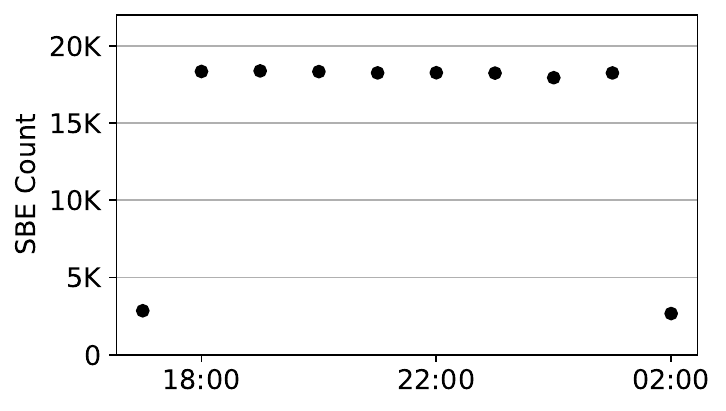}

    (b) Per Hour
	\end{minipage}

    \caption{Example of Delta bursty patterns under different monitoring frequencies from 08/14/2023 to 08/15/2023. Changing monitoring frequency can lead to information change, but no significant information loss is noted.}
    \label{fig:delta-extreme-case}

   \vspace{-\baselineskip}
    
\end{figure}

\begin{figure}[tb]
    \centering
    \begin{minipage}{0.49\columnwidth}
		\footnotesize
  \centering\includegraphics[width=\columnwidth]{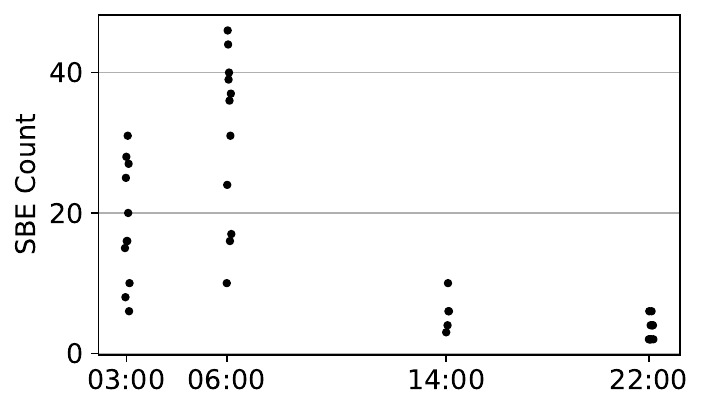}
  
    (a) Per Minute
	\end{minipage}
 	\begin{minipage}{0.49\columnwidth}
  \footnotesize\centering\includegraphics[width=\columnwidth]{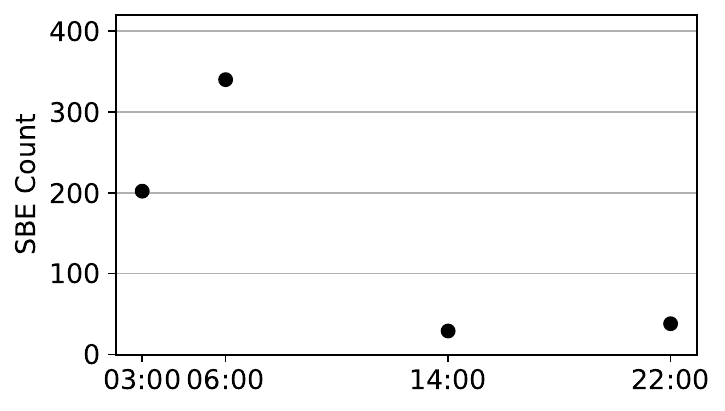}

    (b) Per Hour
	\end{minipage}

    \caption{Another example of Delta bursty patterns shown in different monitoring frequencies on 02/01/2023.}
    \label{fig:delta-extreme-case-2}
\end{figure}

Taking Delta as an example, 
we observe a series of consecutive error occurrence events: there are around 300 SBEs per minute over an eight-hour duration, as shown in \autoref{fig:delta-extreme-case}(a).
We plot the error count in \autoref{fig:delta-extreme-case}(a) using the original error logging frequency of one minute, then present an aggregated hourly error count in \autoref{fig:delta-extreme-case}(b).
With the hour-by-hour error logging, the bursty error pattern is still captured with hourly error counts of around 18,000.
The timespan of the bursty error pattern (up to 8 hours in this case) is sufficiently long, thus the erroneous GPUs are still captured even with the logging frequency of an hour.
Another example is shown in \autoref{fig:delta-extreme-case-2}. Although the hourly error counts capture the errors and indicate some level of burstiness, the bursty error patterns are not as clear as minute-by-minute error logging.
Additionally, error log with higher frequency enables sooner erroneous GPU identification, so that system administrators can take action immediately.

\vspace{0.4cm}
\hspace{-0.15cm}\fbox{\begin{minipage}{0.95\columnwidth}
		\textbf{Take-away Message 2}: Coarser-level error monitoring does not suffer much information loss, yet monitoring errors at a  finer level enables faster responses to errors.  
	\end{minipage}}
\vspace{0.4cm}

\subsubsection{Lessons for GPU Management}

The characterization of spatial and temporal reliability behavior motivates two general key observations that summarize the dynamics of erroneous GPUs: 1) SBEs can be used as an indicator of future DBE occurrence and 2) erroneous GPUs vary over time.
These observations motivate the need for a smart and flexible approach to efficient and low-cost management of supercomputers.
Developing proactive and dynamic GPU monitoring strategies based on erroneous GPU prediction is a potential solution and is subject to our future work.

\vspace{0.4cm}
\hspace{-0.15cm}\fbox{\begin{minipage}{0.95\columnwidth}
		\textbf{Take-away Message 3}: Dynamic GPU monitoring and erroneous GPU prediction strategies are needed for efficient and reliable operation of supercomputers.  
	\end{minipage}}
\vspace{0.4cm}

\subsection{Determining Checkpoint Intervals}

A standard and common practice for handling fail-stop failures is checkpointing, where copies of application state are stored persistently during execution and these copies enable applications to be re-executed from the checkpointed state. 
Checkpointing can require application and system knowledge to accurately capture the application state. Prior studies~\cite{young1974first, daly2006higher} have shown that an optimal checkpoint interval can be determined based on two factors: the time to save a checkpoint into persistent storage, and the mean-time-between-failure of a system.
Taking LFM training as an example, pre-training a model with hundreds of billions of parameters takes thousands of GPUs for months.
One DBE could result in a hardware exception and cause the whole program  
to stop without terminating the batch job (e.g., a SLURM job). 
The allocated computing resources can be wasted if monitoring and restart are not performed in a timely manner.
This problem is exacerbated as state-of-the-art LFM training exceeds $O$(10\textsuperscript{4}) GPUs~\cite{jiang2024megascale}.
The insights from our study can be used as a reference for determining the checkpointing interval of LFM training and general scientific application execution. However, considering the bursty patterns observed in this study, simply computing an optimal but static checkpoint interval might lead to sub-optimal performance and utilization.

An alternative approach is to use dynamic checkpointing based on frequent memory error monitoring.
This dynamic checkpointing strategy should be intelligent enough to tell the period within bursty errors and between bursty errors,
so that the strategy can adjust the checkpoint interval based on the error-free time length as well as upon the observation of SBEs and DBEs. 
In this way, we will be able to minimize the checkpoint overhead while minimizing the failure loss upon uncorrectable memory errors. 

\vspace{0.4cm}
\hspace{-0.15cm}\fbox{\begin{minipage}{0.95\columnwidth}
		\textbf{Take-away Message 4}: The current practice of popular HPC applications for checkpointing is based on a fixed frequency that either fails to meet the reliability requirement or causes high overhead. Dynamic checkpointing is suggested to accommodate bursty error patterns observed in supercomputers.  
	\end{minipage}}
\vspace{0.4cm}

\section{Related Work}

There exist numerous studies of failures for different components in clusters, data centers, and supercomputers~\cite{han2021depth,wang2023understanding,roy2021operating,di2019characterizing,gao2023empirical,das2021systemic,rojas2019analyzing}.
Many works focus on the characterization and analysis of failures in supercomputers~\cite{di2019characterizing,gao2023empirical,das2021systemic,rojas2019analyzing}.
Roy et al.\ perform reliability analysis on the Mira Supercomputer from the perspective of the cooling system~\cite{roy2021operating}. 
Several works delve into silent data corruption (SDC) in data centers, raising awareness within the community of the need for SDC mitigations~\cite{dixit2021silent,wang2023understanding}.
There have been many studies of the impact of memory errors on large machines~\cite{bautista2016unprotected,tiwari2015understanding,nie2016large,nie2017characterizing,ferreira2021understanding,sridharan2013feng,levy2018lessons,beigi2023systematic,zivanovic2019dram,liu2018large}. 
Researchers have examined how correctable errors affect HPC application performance via simulation~\cite{ferreira2021understanding}.
Bautista et al.\ study the raw DRAM error rate without ECC protection on CPU memory~\cite{bautista2016unprotected}.
Feng Shui~\cite{sridharan2013feng} studied the spatial distribution of DRAM and SRAM faults on legacy Cielo and Jaguar supercomputers, confirming that the DRAM and SRAM of Cielo exhibit no aging effects or noticeable increase in the five-year lifetime~\cite{levy2018lessons}.
Recently, Beigi et al. performed a detailed study on DDR4 DRAM faults~\cite{beigi2023systematic}, highlighting concerns regarding multi-bit errors.

On the GPU side, the majority of the studies focus on GPU failures~\cite{taherin2021examining,liu2023predicting,cui2025characterizing} and memory errors~\cite{debardeleben2014gpu,tiwari2015understanding,nie2016large,nie2017characterizing,ostrouchov2020gpu,vazhkudai2017guide,di2014lessons,oles2024understanding}.
Taherin et al. characterize GPU failures and repairs on Tsubame-2 and Tsubame-3 based on system maintenance log, focusing on 
fatal errors such as GPU driver-related problems, kernel panic, and software bugs, but no GPU memory errors were reported by ECC~\cite{taherin2021examining}.
GPU failure logs are used to perform error prediction using machine learning models~\cite{liu2023predicting}.
Most prior GPU memory error studies have been performed on the Titan supercomputer equipped with K20X GPUs~\cite{tiwari2015understanding,nie2016large,nie2017characterizing,ostrouchov2020gpu,vazhkudai2017guide,tiwari2015reliability}.
Periodicity of memory errors is observed in the Titan supercomputer~\cite{nie2016large}.
Di et al. study error behavior of the Blue Waters supercomputer with K20X GPUs~\cite{di2014lessons} but with a focus on CPU-GPU comparison.
The temperature effect is found to be related to GPU memory errors~\cite{nie2016large,nie2017characterizing}.
Debardeleben et al. perform experiments on NVIDIA Tesla M2090 GPUs (using Fermi 2.0 architecture) in Moonlight~\cite{debardeleben2014gpu} and report the rate of DBEs.
These studies mostly focus on earlier generations of GPU architecture.
Most of the machines in these works have been decommissioned at the time of study.

There are two recent studies on the resilience characterization of GPU-accelerated supercomputers.
Oles et al.~\cite{oles2024understanding} perform a case study of GPU memory errors on the Summit supercomputer, equipped with NVIDIA V100 GPUs, focusing on double-bit errors (DBEs) but not single-bit errors (SBEs).
Their analysis explored potential causes of DBEs, such as GPU placement and workload characteristics.
The latest study of GPU reliability is performed on the Delta supercomputer~\cite{cui2025characterizing} where both NVIDIA A100 and H100 GPUs are considered, focusing on GPU hardware failures and interconnect errors but not memory errors.
While we acknowledge that studies based on a single supercomputer can still produce valuable insights, we argue that conclusions drawn from a single system may not generalize to broader GPU deployments, given the varying reliability characteristics across the three supercomputers examined in this work.

In short, our study provides a large-scale, cross-supercomputer analysis of GPU memory errors  on machines currently in production.
As detailed in \autoref{sec:results} and \autoref{sec:implications}, our findings from a cross-supercomputer perspective offer timely and practical insights into the reliability of modern GPU-based systems, filling an important gap in the existing literature.

\section{Conclusions and Future Work}

We have presented an in-depth quantitative error analysis of memory errors in Ampere GPUs across three heavily utilized, in-production supercomputers of varying scales: Delta~\cite{delta}, Polaris~\cite{polaris}, and Perlmutter~\cite{perlmutter}.
We collect and analyze GPU memory ECC (Error Correction Codes) error logs, resulting in a total of 67.77  million GPU device-hours of data covering 10,693 NVIDIA Ampere GPUs.

We quantitatively measure the error rate and Mean-Time-Between-Errors (MTBE) in these clusters.
The burstiness, periodicity, spatial and temporal relation of errors, and environmental factors in these clusters are also explored in detail.
The key observations include 1) Bursty error patterns have a significant impact on the characteristics of error rate and MTBE;
2) Cluster scale also affects MTBE but the relationship is not linear;
3) Errors have spatial correlation but exhibit no periodical
patterns; 
4) No strong correlation of errors is observed with environmental factors such as temperature, power, and GPU utilization; and
5) GPU errors are spatially and temporally correlated.

Furthermore, we summarize the lessons learned from this study. We also compare the
Ampere GPU error statistics with previous generations to
observe the trend of GPU reliability and understand the nature of GPU errors.
We discuss the opportunities for efficient
machine health management and large-scale application checkpointing.
We bring up the possibility of dynamic checkpointing strategies informed by memory error monitoring.
Our observations and analyses motivate the future studies to explore the possibility of designing error prediction models and dynamic checkpointing  algorithms based.

\section*{Acknowledgments}

This material is based upon work supported by the National 
Science Foundation (NSF) grants (\#2402940 and \#2410856) and the Commonwealth Cyber Initiative (CCI) grant (\#HC‐3Q24‐047). The platforms used for evaluation in this work are
supported by the U.S. DOE Office of Science, Office of Advanced Scientific Computing Research, under award 66150:
"CENATE - Center for Advanced Architecture Evaluation".
The Pacific Northwest National Laboratory is operated by
Battelle for the U.S. Department of Energy under Contract
DE-AC05-76RL01830.
This project is supported by resources provided by the Office of Research Computing at George Mason University (URL: https://orc.gmu.edu) and funded in part by grants from the National Science Foundation (Award Number 2018631).
This research used both the DeltaAI advanced computing and data resource, which is supported by the National Science Foundation (award OAC 2320345) and the State of Illinois, and the Delta advanced computing and data resource which is supported by the National Science Foundation (award OAC 2005572) and the State of Illinois.. Delta and DeltaAI are joint efforts of the University of Illinois Urbana-Champaign and its National Center for Supercomputing Applications.
This research used resources of the Argonne Leadership Computing Facility, a U.S. Department of Energy (DOE) Office of Science user facility at Argonne National Laboratory and is based on research supported by the U.S. DOE Office of Science-Advanced Scientific Computing Research Program, under Contract No. DE-AC02-06CH11357.
This research used resources of the National Energy Research Scientific Computing Center (NERSC), a Department
 of Energy Office of Science User Facility using NERSC
 award DDR-ERCAP0032037.

\balance
\bibliographystyle{ieeetr}
\bibliography{GPUError}

\begin{thebibliography}{10}

\bibitem{chen2021gpu}
Y.~Chen, W.~Li, R.~Fan, and X.~Liu, ``Gpu optimization for high-quality kinetic fluid simulation,'' {\em IEEE Transactions on Visualization and Computer Graphics}, vol.~28, no.~9, pp.~3235--3251, 2021.

\bibitem{liu2004real}
Y.~Liu, X.~Liu, and E.~Wu, ``Real-time 3d fluid simulation on gpu with complex obstacles,'' in {\em 12th Pacific Conference on Computer Graphics and Applications, 2004. PG 2004. Proceedings.}, pp.~247--256, IEEE, 2004.

\bibitem{glaser2015strong}
J.~Glaser, T.~D. Nguyen, J.~A. Anderson, P.~Lui, F.~Spiga, J.~A. Millan, D.~C. Morse, and S.~C. Glotzer, ``Strong scaling of general-purpose molecular dynamics simulations on gpus,'' {\em Computer Physics Communications}, vol.~192, pp.~97--107, 2015.

\bibitem{boku2024improving}
T.~Boku, M.~Sugita, R.~Kobayashi, S.~Furuya, T.~Fujie, M.~Ohue, and Y.~Akiyama, ``Improving performance on replica-exchange molecular dynamics simulations by optimizing gpu core utilization,'' in {\em Proceedings of the 53rd International Conference on Parallel Processing}, pp.~1082--1091, 2024.

\bibitem{radford2018improving}
A.~Radford, K.~Narasimhan, T.~Salimans, and I.~Sutskever, ``Improving language understanding with unsupervised learning,'' 2018.
\newblock Technical report, OpenAI. \url{https://openai.com/index/language-unsupervised/}.

\bibitem{zhang2022opt}
S.~Zhang, S.~Roller, N.~Goyal, M.~Artetxe, M.~Chen, S.~Chen, C.~Dewan, M.~Diab, X.~Li, X.~V. Lin, T.~Mihaylov, M.~Ott, S.~Shleifer, K.~Shuster, D.~Simig, P.~S. Koura, A.~Sridhar, T.~Wang, and L.~Zettlemoyer, ``{OPT}: Open pre-trained transformer language models,'' {\em Preprint arXiv:2205.01068}, 2022.

\bibitem{zvyagin2022genslm}
M.~Zvyagin, A.~Brace, K.~Hippe, Y.~Deng, B.~Zhang, C.~Orozco~Bohorquez, A.~Clyde, B.~Kale, D.~Perez-Rivera, H.~Ma, C.~M. Mann, M.~Irvin, J.~G. Pauloski, L.~Ward, V.~Hayot, M.~Emani, S.~Foreman, Z.~Xie, D.~Lin, M.~Shukla, W.~Nie, J.~Romero, C.~Dallago, A.~Vahdat, C.~Xiao, T.~Gibbs, I.~Foster, J.~J. Davis, M.~E. Papka, T.~Brettin, R.~Stevens, A.~Anandkumar, V.~Vishwanath, and A.~Ramanathan, ``{GenSLMs: Genome-scale language models reveal SARS-CoV-2 evolutionary dynamics},'' {\em The International Journal of High Performance Computing Applications}, vol.~37, no.~6, pp.~683--705, 2023.

\bibitem{DBLP:conf/dsn/FratinOLSRR18}
V.~Fratin, D.~A.~G. de~Oliveira, C.~B. Lunardi, F.~Santos, G.~Rodrigues, and P.~Rech, ``Code-dependent and architecture-dependent reliability behaviors,'' in {\em {DSN}}, pp.~13--26, 2018.

\bibitem{killi}
S.~Ganapathy, J.~Kalamatianos, B.~M. Beckmann, S.~Raasch, and L.~G. Szafaryn, ``Killi: Runtime fault classification to deploy low voltage caches without {MBIST},'' in {\em 25th IEEE International Symposium on High Performance Computer Architecture}, pp.~304--316, {IEEE}, 2019.

\bibitem{sullivan2021characterizing}
M.~B. Sullivan, N.~Saxena, M.~O'Connor, D.~Lee, P.~Racunas, S.~Hukerikar, T.~Tsai, S.~K.~S. Hari, and S.~W. Keckler, ``Characterizing and mitigating soft errors in {GPU} dram,'' in {\em MICRO-54: 54th Annual IEEE/ACM International Symposium on Microarchitecture}, pp.~641--653, 2021.

\bibitem{opt175logbook}
{MetaSeq}, ``{OPT-175B Baselines} logbook,'' 2022.
\newblock \url{https://github.com/facebookresearch/metaseq/blob/main/projects/OPT/chronicles/OPT175B_Logbook.pdf}.

\bibitem{dixit2021silent}
H.~D. Dixit, S.~Pendharkar, M.~Beadon, C.~Mason, T.~Chakravarthy, B.~Muthiah, and S.~Sankar, ``Silent data corruptions at scale,'' {\em Preprint arXiv:2102.11245}, 2021.

\bibitem{he2023understanding}
Y.~He, M.~Hutton, S.~Chan, R.~De~Gruijl, R.~Govindaraju, N.~Patil, and Y.~Li, ``Understanding and mitigating hardware failures in deep learning training systems,'' in {\em 50th Annual International Symposium on Computer Architecture}, pp.~1--16, 2023.

\bibitem{zhang2019quantifying}
Z.~Zhang, L.~Huang, R.~Huang, W.~Xu, and D.~S. Katz, ``Quantifying the impact of memory errors in deep learning,'' in {\em IEEE International Conference on Cluster Computing}, pp.~1--12, IEEE, 2019.

\bibitem{li2017understanding}
G.~Li, S.~K.~S. Hari, M.~Sullivan, T.~Tsai, K.~Pattabiraman, J.~Emer, and S.~W. Keckler, ``Understanding error propagation in deep learning neural network {(DNN)} accelerators and applications,'' in {\em International Conference for High Performance Computing, Networking, Storage and Analysis}, pp.~1--12, 2017.

\bibitem{tr176Blogbook}
{BigScience}, ``{TR11-176B Training Logbook},'' 2022.
\newblock \url{https://github.com/bigscience-workshop/bigscience/blob/master/train/tr11-176B-ml/chronicles.md}.

\bibitem{levy2018lessons}
S.~Levy, K.~B. Ferreira, N.~DeBardeleben, T.~Siddiqua, V.~Sridharan, and E.~Baseman, ``Lessons learned from memory errors observed over the lifetime of {C}ielo,'' in {\em SC18: International Conference for High Performance Computing, Networking, Storage and Analysis}, pp.~554--565, IEEE, 2018.

\bibitem{bautista2016unprotected}
L.~Bautista-Gomez, F.~Zyulkyarov, O.~Unsal, and S.~McIntosh-Smith, ``Unprotected computing: A large-scale study of {DRAM} raw error rate on a supercomputer,'' in {\em SC'16: International Conference for High Performance Computing, Networking, Storage and Analysis}, pp.~645--655, IEEE, 2016.

\bibitem{ferreira2021understanding}
K.~B. Ferreira, S.~Levy, V.~Kuhns, N.~DeBardeleben, and S.~Blanchard, ``Understanding the effects of {DRAM} correctable error logging at scale,'' in {\em IEEE International Conference on Cluster Computing}, pp.~421--432, IEEE, 2021.

\bibitem{sridharan2013feng}
V.~Sridharan, J.~Stearley, N.~DeBardeleben, S.~Blanchard, and S.~Gurumurthi, ``Feng {S}hui of supercomputer memory: Positional effects in {DRAM and SRAM} faults,'' in {\em International Conference on High Performance Computing, Networking, Storage and Analysis}, pp.~1--11, 2013.

\bibitem{beigi2023systematic}
M.~V. Beigi, Y.~Cao, S.~Gurumurthi, C.~Recchia, A.~Walton, and V.~Sridharan, ``A systematic study of {DDR4 DRAM} faults in the field,'' in {\em IEEE International Symposium on High-Performance Computer Architecture}, pp.~991--1002, IEEE, 2023.

\bibitem{di2014lessons}
C.~Di~Martino, Z.~Kalbarczyk, R.~K. Iyer, F.~Baccanico, J.~Fullop, and W.~Kramer, ``Lessons learned from the analysis of system failures at petascale: The case of {Blue Waters},'' in {\em 44th Annual IEEE/IFIP International Conference on Dependable Systems and Networks}, pp.~610--621, IEEE, 2014.

\bibitem{debardeleben2014gpu}
N.~DeBardeleben, S.~Blanchard, L.~Monroe, P.~Romero, D.~Grunau, C.~Idler, and C.~Wright, ``{GPU behavior on a large HPC} cluster,'' in {\em Euro-Par Workshops 2013}, pp.~680--689, Springer, 2014.

\bibitem{tiwari2015understanding}
D.~Tiwari, S.~Gupta, J.~Rogers, D.~Maxwell, P.~Rech, S.~Vazhkudai, D.~Oliveira, D.~Londo, N.~DeBardeleben, P.~Navaux, L.~Carrot, and A.~Bland, ``Understanding {GPU} errors on large-scale {HPC} systems and the implications for system design and operation,'' in {\em IEEE 21st International Symposium on High Performance Computer Architecture}, pp.~331--342, IEEE, 2015.

\bibitem{nie2016large}
B.~Nie, D.~Tiwari, S.~Gupta, E.~Smirni, and J.~H. Rogers, ``A large-scale study of soft-errors on {GPUs} in the field,'' in {\em IEEE International Symposium on High Performance Computer Architecture}, pp.~519--530, IEEE, 2016.

\bibitem{nie2017characterizing}
B.~Nie, J.~Xue, S.~Gupta, C.~Engelmann, E.~Smirni, and D.~Tiwari, ``Characterizing temperature, power, and soft-error behaviors in data center systems: Insights, challenges, and opportunities,'' in {\em IEEE 25th International Symposium on Modeling, Analysis, and Simulation of Computer and Telecommunication Systems}, pp.~22--31, IEEE, 2017.

\bibitem{ostrouchov2020gpu}
G.~Ostrouchov, D.~Maxwell, R.~A. Ashraf, C.~Engelmann, M.~Shankar, and J.~H. Rogers, ``{GPU} lifetimes on {T}itan supercomputer: Survival analysis and reliability,'' in {\em SC20: International Conference for High Performance Computing, Networking, Storage and Analysis}, pp.~1--14, IEEE, 2020.

\bibitem{oles2024understanding}
V.~Oles, A.~Schmedding, G.~Ostrouchov, W.~Shin, E.~Smirni, and C.~Engelmann, ``Understanding gpu memory corruption at extreme scale: The summit case study,'' in {\em Proceedings of the 38th ACM International Conference on Supercomputing}, pp.~188--200, 2024.

\bibitem{cui2025characterizing}
S.~Cui, A.~Patke, Z.~Chen, A.~Ranjan, H.~Nguyen, P.~Cao, S.~Jha, B.~Bode, G.~Bauer, C.~Narayanaswami, {\em et~al.}, ``Characterizing gpu resilience and impact on ai/hpc systems,'' {\em arXiv preprint arXiv:2503.11901}, 2025.

\bibitem{delta}
{National Center for Supercomputing Applications}, ``{Delta}.''
\newblock \url{https://www.ncsa.illinois.edu/research/project-highlights/delta/}.

\bibitem{polaris}
{Argonne National Laboratory}, ``{Polaris}.''
\newblock \url{https://www.alcf.anl.gov/polaris}.

\bibitem{perlmutter}
{National Energy Research Scientific Computing Center}, ``{Perlmutter}.''
\newblock \url{https://docs.nersc.gov/systems/perlmutter/architecture/}.

\bibitem{nvidia2020ampere}
{NVIDIA Corporation}, ``{NVIDIA Ampere Architecture Whitepaper}.''
\newblock \url{https://www.nvidia.com/content/PDF/nvidia-ampere-ga-102-gpu-architecture-whitepaper-v2.pdf}.

\bibitem{dcgm}
{NVIDIA Corporation}, ``{NVIDIA Data Center GPU Manager (DCGM)}.''
\newblock \url{https://docs.nvidia.com/data-center-gpu-manager-dcgm/index.html}.

\bibitem{top500}
Top500, ``{NOVEMBER 2023 Top500 Supercomputers}.''
\newblock \url{https://www.top500.org/lists/top500/2023/11/}.

\bibitem{leemis2006discrete}
L.~M. Leemis and S.~K. Park, {\em Discrete-event simulation: A first course}.
\newblock Pearson Prentice Hall Upper Saddle River, 2006.

\bibitem{young1974first}
J.~W. Young, ``A first order approximation to the optimum checkpoint interval,'' {\em Communications of the ACM}, vol.~17, no.~9, pp.~530--531, 1974.

\bibitem{daly2006higher}
J.~T. Daly, ``A higher order estimate of the optimum checkpoint interval for restart dumps,'' {\em Future Generation Computer Systems}, vol.~22, no.~3, pp.~303--312, 2006.

\bibitem{jiang2024megascale}
Z.~Jiang, H.~Lin, Y.~Zhong, Q.~Huang, Y.~Chen, Z.~Zhang, Y.~Peng, X.~Li, C.~Xie, S.~Nong, Y.~Jia, S.~He, H.~Chen, Z.~Bai, Q.~Hou, S.~Yan, D.~Zhou, Y.~Sheng, Z.~Jiang, H.~Xu, H.~Wei, Z.~Zhang, P.~Nie, L.~Zou, S.~Zhao, L.~Xiang, Z.~Liu, Z.~Li, X.~Jia, J.~Ye, X.~Jin, and X.~Liu, ``Mega{S}cale: Scaling large language model training to more than 10,000 {GPUs},'' {\em Preprint arXiv:2402.15627}, 2024.

\bibitem{han2021depth}
S.~Han, P.~P. Lee, F.~Xu, Y.~Liu, C.~He, and J.~Liu, ``An in-depth study of correlated failures in production {SSD-Based} data centers,'' in {\em 19th USENIX Conference on File and Storage Technologies}, pp.~417--429, 2021.

\bibitem{wang2023understanding}
S.~Wang, G.~Zhang, J.~Wei, Y.~Wang, J.~Wu, and Q.~Luo, ``Understanding silent data corruptions in a large production {CPU} population,'' in {\em 29th Symposium on Operating Systems Principles}, pp.~216--230, 2023.

\bibitem{roy2021operating}
R.~B. Roy, T.~Patel, R.~Kettimuthu, W.~Allcock, P.~Rich, A.~Scovel, and D.~Tiwari, ``Operating liquid-cooled large-scale systems: Long-term monitoring, reliability analysis, and efficiency measures,'' in {\em IEEE International Symposium on High-Performance Computer Architecture}, pp.~881--893, IEEE, 2021.

\bibitem{di2019characterizing}
S.~Di, H.~Guo, E.~Pershey, M.~Snir, and F.~Cappello, ``Characterizing and understanding {HPC} job failures over the 2{K}-day life of {IBM} {BlueGene/A} system,'' in {\em 49th Annual IEEE/IFIP International Conference on Dependable Systems and Networks}, pp.~473--484, IEEE, 2019.

\bibitem{gao2023empirical}
Y.~Gao, X.~Shi, H.~Lin, H.~Zhang, H.~Wu, R.~Li, and M.~Yang, ``An empirical study on quality issues of deep learning platform,'' in {\em IEEE/ACM 45th International Conference on Software Engineering: Software Engineering in Practice}, pp.~455--466, IEEE, 2023.

\bibitem{das2021systemic}
A.~Das, F.~Mueller, and B.~Rountree, ``Systemic assessment of node failures in {HPC} production platforms,'' in {\em IEEE International Parallel and Distributed Processing Symposium}, pp.~267--276, IEEE, 2021.

\bibitem{rojas2019analyzing}
E.~Rojas, E.~Meneses, T.~Jones, and D.~Maxwell, ``Analyzing a five-year failure record of a leadership-class supercomputer,'' in {\em 31st International Symposium on Computer Architecture and High Performance Computing}, pp.~196--203, IEEE, 2019.

\bibitem{zivanovic2019dram}
D.~Zivanovic, P.~E. Dokht, S.~Mor{\'e}, J.~Bartolome, P.~M. Carpenter, P.~Radojkovi{\'c}, and E.~Ayguad{\'e}, ``{DRAM} errors in the field: A statistical approach,'' in {\em International Symposium on Memory Systems}, pp.~69--84, 2019.

\bibitem{liu2018large}
R.-T. Liu and Z.-N. Chen, ``A large-scale study of failures on petascale supercomputers,'' {\em Journal of Computer Science and Technology}, vol.~33, no.~1, pp.~24--41, 2018.

\bibitem{taherin2021examining}
A.~Taherin, T.~Patel, G.~Georgakoudis, I.~Laguna, and D.~Tiwari, ``Examining failures and repairs on supercomputers with multi-{GPU} compute nodes,'' in {\em 51st Annual IEEE/IFIP International Conference on Dependable Systems and Networks}, pp.~305--313, IEEE, 2021.

\bibitem{liu2023predicting}
H.~Liu, Z.~Li, C.~Tan, R.~Yang, G.~Cao, Z.~Liu, and C.~Guo, ``Predicting {GPU} failures with high precision under deep learning workloads,'' in {\em 16th ACM International Conference on Systems and Storage}, pp.~124--135, 2023.

\bibitem{vazhkudai2017guide}
S.~S. Vazhkudai, R.~Miller, D.~Tiwari, C.~Zimmer, F.~Wang, S.~Oral, R.~Gunasekaran, and D.~Steinert, ``{GUIDE}: a scalable information directory service to collect, federate, and analyze logs for operational insights into a leadership {HPC} facility,'' in {\em International Conference for High Performance Computing, Networking, Storage and Analysis}, pp.~1--12, 2017.

\bibitem{tiwari2015reliability}
D.~Tiwari, S.~Gupta, G.~Gallarno, J.~Rogers, and D.~Maxwell, ``Reliability lessons learned from {GPU} experience with the {T}itan supercomputer at {Oak ridge Leadership Computing Facility},'' in {\em International Conference for High Performance Computing, Networking, Storage and Analysis}, pp.~1--12, 2015.

\end{thebibliography}

\end{document}